\documentclass{aa}
\usepackage{hyperref}
\usepackage[varg]{txfonts}
\usepackage{soul}
\usepackage{xcolor}
\begin{document}

\title{The mass distribution in the outskirts of clusters of galaxies as \\ a probe of the theory of gravity}

\author{Michele Pizzardo \inst{\ref{1},\ref{2},\ref{3}}
\and Antonaldo Diaferio \inst{\ref{2},\ref{3}}
\and Kenneth J. Rines \inst{\ref{4}}
}
\institute{\label{1}Department of Astronomy and Physics, Saint Mary's University, 923 Robie Street, Halifax, NS-B3H3C3, Canada
\and \label{2}Dipartimento di Fisica, Universit\`a di Torino, via P. Giuria 1,  I-10125 Torino, Italy
\and \label{3}Istituto Nazionale di Fisica Nucleare (INFN), Sezione di Torino, via P. Giuria 1,  I-10125 Torino, Italy
\and \label{4}Department of Physics and Astronomy, Western Washington University, Bellingham, WA-98225, USA
} 

\date{Received date / Accepted date}

\abstract
{We show that $\varsigma$,  the radial location of the minimum in the differential radial mass profile $M^\prime(r)$  of a galaxy cluster, can probe the theory of gravity. We derived $M^\prime(r)$ of the dark matter halos of galaxy clusters from N-body cosmological simulations that implement two different theories of gravity: standard gravity in the $\Lambda$CDM model, and $f(R)$. 
We extracted 49169 dark matter halos in 11 redshift bins in the range $0\leq z\leq 1$ and in three different mass bins in the range $0.9<M_{200c}/10^{14}h^{-1}$M$_\odot<11$. We investigated the correlation of $\varsigma$ with the redshift and the mass accretion rate (MAR) of the halos. 
We show that $\varsigma$ decreases from $\sim 3R_{200c}$ to $\sim 2R_{200c}$ when $z$ increases from 0 to $1$ in the $\Lambda$CDM model. At $z\sim 0.1$, $\varsigma$ decreases from $2.8R_{200c}$ to $\sim 2.5R_{200c}$ when the MAR increases from $\sim 10^4h^{-1}$M$_\odot$~yr$^{-1}$ to  $\sim 2\times 10^5h^{-1}$M$_\odot$~yr$^{-1}$. In the $f(R)$ model, $\varsigma$ is $\sim 15$\% larger than in $\Lambda$CDM. 
The median test shows that for samples of $\gtrsim 400$ dark matter halos at $z\leq 0.8$, $\varsigma$ is able to distinguish between the two theories of gravity with a $p$-value $\lesssim 10^{-5}$.
Upcoming advanced spectroscopic and photometric programs will allow a robust estimation of the mass profile of enormous samples of clusters up to large clustercentric distances. These samples will allow us to statistically exploit $\varsigma$ as probe of the theory of gravity, which complements other large-scale probes.
} 


\maketitle

\section{Introduction}

According to the standard $\Lambda$CDM model, the gravitational dynamics of the Universe and its content is described by General Relativity in a flat space-time with a positive cosmological constant, $\Lambda$. The matter content of the Universe mainly consists of cold dark matter (CDM), and baryons are a minor component. In this model, cosmological structures form hierarchically through consecutive mergers of smaller halos into larger halos \citep[e.g.,][]{press1974formation,white1978,laceyCole93,bower1991,sheth2002,zhang2008,desimone2011,corasaniti2011,achitouv2014,musso2018}.             

For a long time, the outskirts of clusters of galaxies have been regarded as a potentially powerful probe of the model of structure formation and evolution \citep[e.g.,][]{diaferio2004outskirts,reiprich2013outskirts,Diemer2014,lau2015mass,walker2019physics,rost2021threehundred}. At radii larger than the virial radius, this region is linked to the cosmic large-scale structure, which is sensitive to the model of gravity and to the nature of dark matter. Furthermore, unlike the central regions of clusters, comparisons of the dynamical properties of these regions with simulations are more robust because the relevant physical processes are dominated by gravity and not by baryonic physics \citep[e.g.,][]{diemand2004,hellwing2016,armitage2018,velliscig2014,vandaalen2014,shirasaki2018}. 

The most widely investigated feature of cluster outskirts is the splashback radius, $R_{\rm spl}$, which roughly corresponds to the first apocenter of the orbits of recently accreted material \citep{Adhikari2014}. This radius has an important physical meaning: It separates particles that are orbiting in the gravitational potential of the halo from matter falling onto it \citep{Diemer2014,Adhikari2014,More2015,Diemer20,Diemer21}. For this reason, \citet{More2015} proposed it as the physical boundary of clusters. \citet{Diemer2014} and \citet{More2015} showed that this feature can be detected as a sudden drop in the logarithmic slope of the halo density profile. These works employed N-body simulations, and they showed that $R_{\rm spl}$ is usually located at $\sim 2R_{200c}$\footnote{In the following, $R_{200c}$ will denote the radius within which the mean density equals 200 times the critical density $\rho_c$. It is a common proxy of the virial radius \citep{bryan1998}.} and that it decreases with increasing mass accretion rate (MAR) and redshift.

The splashback radius is a genuine prediction of the hierachical model of structure formation. It is therefore expected to be sensitive to the model of gravity and the nature of dark matter and dark energy \citep{Diemer2014,Adhikari2014,adhikari2018splashback,contigiani2019,Banerjee20,Diemer20,Umetsu20essay}. \citet{adhikari2018splashback} proposed the exploitation of $R_{\rm spl}$ as a probe of cosmic expansion and the model of gravity for the first time. They used analytical models and N-body simulations to study the impact of dark energy and modifications in gravity on the value of $R_{\rm spl}$. They showed that this feature is sensitive both to the $w$ parameter of the equation of state of the dark energy fluid and to two screened modified gravity models, $f(R)$ and nDGP. They pointed out that the outskirts of galaxy clusters are the transition region between the screened and unscreened regimes of modified gravity models, and thus they are the ideal region in which to detect signatures of modified gravity. 

In the data, several works detected a change in the slope of the mass density profile at a radius compatible with the expected $R_{\rm spl}$ \citep[e.g.,][]{More16,Baxter17,umetsu2017lensing,Chang_2018,Zurcher19,Shin19,Murata20,Bianconi21,Gonzalez21}. 
However, although these results are encouraging, observational estimates are affected by both projection effects and by the fact that we observe galaxies rather than dark matter particles. Dark matter particles and dark matter subhalos, with which galaxies might be associated, do not necessarily share the same apocenter \citep{Diemer2017sparta2}.  Recent investigations based on hydrodynamical simulations showed that $R_{\rm spl}$ derived from galaxies or gas particles can substantially differ from $R_{\rm spl}$ derived from dark matter \citep{Baxter21,Deason21,oNeil21,Dacunha22,oneil22}. These discrepancies behave in a complex way that depends on the halo mass and selection criteria.
Therefore, further observational and methodological efforts are required.

High-precision measurements require high-quality spectroscopic and wide-field surveys. These will be available in the near future by ongoing or planned experiments; in addition, an improved understanding of the systematics that affect the methods that are used to estimate the mass profile of clusters is necessary. Currently, weak gravitational lensing \citep{bartelmann2010gravitational,Umetsu20essay} and the caustic technique \citep{Diaferio99,Serra2011} are the only methods that estimate the cluster mass profile at large cluster-centric distances without relying on dynamical equilibrium. Weak-lensing is steadily improving with the acquisition of large combined spectroscopic and photometric datasets and methodological advancements \citep{Umetsu13,dellantonio20,Umetsu20essay}. Synergies with the caustic technique \citep{Diaferio99,Serra2011} promise unbiased and accurate mass profiles that extend to large radii. 

On the theoretical side, we need to thoroughly characterize the dynamics of clusters in their outer region. This plan requires investigating other characteristic radii of cluster outskirts in addition to $R_{\rm spl}$. Beyond $R_{\rm spl}$, clusters present an extended region in which matter is falling onto them. The radius at which the radial velocity profile has its minimum traces the region of clusters in which the dynamics is a genuine infall \citep{deBoni2016,valles20,pizzardo2020,Pizzardo23mar}. Characteristic radii also arise from the halo-matter bias \citep{fong21,Garcia21}. These radii depend on the cluster mass and redshift. Because these radii are tightly connected with the cluster MAR, they are probably extremely sensitive to the gravity model. Studies aiming at the quantitative determination of the dependence of these radii on the gravity model are vital for the exploitation of cluster outskirts as a cosmological probe.

Here, we focus on a radius that is different from both $R_{\rm spl}$ and the radius of the minimum of the radial velocity profile. We consider the radial location, $\varsigma$, of the minimum of the radial differential mass profile of galaxy clusters, $M^\prime(r)$. We investigate whether $\varsigma$ can distinguish between standard gravity and $f(R)$  gravity based on galaxy cluster halos in wide redshift and mass ranges. 
The radius $\varsigma$ complements $R_{\rm spl}$ as a probe of the external region of clusters of galaxies: $\varsigma$ reduces the large errors in $R_{\rm spl}$ that originate from the logarithmic slope of the density profile, and it can thus in principle be located in individual systems. We prove that although $ \varsigma$ and  $R_{\rm spl}$ are not equivalent, they share similar trends with redshift, mass, and MAR.

The paper is organized as follows. In Sect. \ref{pres} we present the simulations (Sect. \ref{sim}) and the samples of simulated halos (Sect. \ref{numRes1}). In Sect. \ref{numRes} we present our theoretical results: in Sect. \ref{numRes2} we compare $\varsigma$ with $R_{\rm spl}$, and in Sect. \ref{numRes3} we study whether $\varsigma$ can distinguish between two different gravity models. We discuss our results in Sect. \ref{discussion} and conclude in Sect. \ref{conclusion}.

\section{Simulations and samples of halos}\label{pres}

\subsection{DUSTGRAIN simulations}\label{sim}

We extracted our samples of simulated halos from DUSTGRAIN \citep{giocoli2018weak}, which is a suite of N-body cosmological simulations implementing the standard $\Lambda$CDM model and various extensions of General Relativity in the form of $f(R)$. The simulations might include 
a non-negligible fraction of massive neutrinos. Appendix \ref{append} reports a basic review of the $f(R)$ model considered in these simulations. 
For our purposes, we used only the $\Lambda$CDM and the $fR4$ runs; the latter is an $f(R)$ run with scalaron $f_{R0}=-1\cdot 10^{-4}$ and without massive neutrinos ($\Omega_{\nu}=0$).

The DUSTGRAIN simulations were normalized at the cosmic microwave background (CMB) epoch with cosmological parameters consistent with the {\it{Planck 2015}} constraints \citep{ade2016planck-xiii}. The cosmological matter density is $\Omega_{M}\equiv\Omega_{CDM}+\Omega_{b}+\Omega_{\nu}=0.31345$, where $\Omega_{CDM}$, $\Omega_b$, and $\Omega_{\nu}$ are the current CDM, baryonic, and neutrino density parameters, respectively. The DUSTGRAIN simulations had $\Omega_{b}=0$. Furthermore, they had a cosmological constant $\Omega_{\Lambda}=0.68655$, a Hubble constant $H_0=100h$~km~s$^{-1}$~Mpc$^{-1}$, with $h=0.6731$, an initial perturbation amplitude $\mathcal{A}_s=2.199\times 10^{-9}$, and a power spectrum index $n_s=0.9658$. The normalization of the power spectrum of the perturbations at the CMB epoch implies that the value of the power spectrum normalization at $z=0$, $\sigma_8$, generally differs among the different simulations. In the two runs relevant for this work, $\Lambda$CDM and $fR4$, $\sigma_8=0.847$ and $\sigma_8=0.967$, respectively.\footnote{The different values of $\sigma_8$ in the two simulations do not affect our results, as discussed in Sect. \ref{discussion}.} The box size was $750\, h^{-1} {\rm{Mpc}} $ on a side in comoving coordinates. The simulation contained $768^3$ collisionless dark matter particles with mass $m_{\rm DM} = 8.1 \times 10^{10} {h^{-1}\rm{M_{\odot}}} $ each. Because hydrodynamics is absent, the $\Lambda$CDM and the $fR4$ runs are two standard collisionless $N$-body simulations with $\Omega_M=\Omega_{CDM}=0.31345$ and $\Omega_b=\Omega_\nu = 0$.

The CDM halos were identified with a friends-of-friends (FoF) algorithm \citep{Davis85} with linking length $\lambda=0.16\,\bar{d}$, where $\bar{d}$ is the mean Lagrangian interparticle separation. The SUBFIND algorithm \citep{subfind2001} was run on the FoF groups to identify their gravitationally bound substructures. A synthetic cluster corresponds to the most massive detected substructure, whose most tightly bound particle is the cluster center. 

\subsection{Samples of simulated halos} \label{numRes1}

For the theoretical investigations presented in this work, we considered the three-dimensional synthetic clusters of DUSTGRAIN (Sect. \ref{sim}) in 11 redshift bins in the range $0\leq z\leq 1$. Table \ref{samples} lists the main properties of the cluster samples at each redshift: the number of clusters, the median mass, and the median MAR with the $68$th percentile ranges of these quantities.

The clusters were separated into  low-, intermediate-, and high-mass samples. The low-mass samples contained all the simulated halos with mass $M_{200c}\in (0.9,1.1)\cdot 10^{14}\, h^{-1}$M$_\odot$, and the intermediate- and high-mass samples contained all the halos with mass $M_{200c}\in (4.0,6.0)\cdot 10^{14}\, h^{-1}$M$_\odot$ and $M_{200c}\in (9.0,11.0)\cdot 10^{14}\, h^{-1}$M$_\odot$, respectively. The high-mass samples are missing at $z>0.3$ because the number of  halos in this mass range is  $\sim 0-10$ and these small samples would not provide statistically robust results.  

We had 26 independent samples of three-dimensional halos for each of the two models we considered: 11 samples of low-mass (L) halos, 11 samples of intermediate-mass (I) halos, and 4 samples of high-mass (H) halos. In total, the 26 $\Lambda$CDM ($fR4$) halo samples are provided with  the six phase-space coordinates of the dark matter particles in 20887 (28282) synthetic clusters; the higher number of halos for $fR4$ is the result of the higher $\sigma_8$ of this model. 

As an effect of the cluster mass function, the low-mass samples have $\sim 2000$ halos at each redshift (1736 and 2303, averaging over all redshifts, for $\Lambda$CDM and $fR4$, respectively), the intermediate-mass samples have $\sim 200$ halos (154 and 251, averaging over all redshifts, for $\Lambda$CDM and $fR4$, respectively), and the high-mass samples have $\sim 35$ halos (25 and 46, averaging over all redshifts, for $\Lambda$CDM and $fR4$, respectively). The redshift dependence of the mass function has clear consequences for the size of the samples in the same mass bin but different redshift: for the low-mass samples, the number of halos at $z=0.00$ is $\sim 3-4$ times higher than at $z=1.00$; for the intermediate-mass samples, this ratio is $\sim 28-38$. 

\begin{table*}[h]
\begin{center}
\caption{ Samples of simulated clusters.}\label{samples}
\begin{tabular}{c|ccc|ccc|ccc}
\hline
\hline
$z$ & \multicolumn{3}{c|}{$\Lambda$CDM-L} & \multicolumn{3}{c|}{$\Lambda$CDM-I} & \multicolumn{3}{c}{$\Lambda$CDM-H} \\
\hline
    & size & $ M_{200c}$ & MAR $(\Gamma_{\rm dyn})$ & size & $M_{200c}$ & MAR $(\Gamma_{\rm dyn})$ & size & $M_{200c}$ & MAR $(\Gamma_{\rm dyn})$ \\
       &    & {\small $[10^{14}\, h^{-1}$M$_\odot]$} & {\small $[10^{3}\, h^{-1}$M$_\odot]$~yr$^{-1}$} & & {\small $[10^{14}\, h^{-1}$M$_\odot]$} & {\small $[10^{3}\, h^{-1}$M$_\odot]$~yr$^{-1}$} & & {\small $[10^{14}\, h^{-1}$M$_\odot]$} & {\small $[10^{3}\, h^{-1}$M$_\odot]$~yr$^{-1}$} \\

\hline
0.00    &       2755    &       $0.992_{-0.065}^{+0.073}$       &       $11.9_{-4.6}^{+7.6} \, (1.73)$      &       407     &       $4.64_{-0.46}^{+0.80}$  &       $77_{-28}^{+52} \,(2.38)$       &       41      &       $9.74_{-0.45}^{+0.78}$ & $174_{-77}^{+102} \,(2.58)$       \\
0.09    &       2656    &       $0.989_{-0.062}^{+0.071}$       &       $14.1_{-5.4}^{+10.3} \, (1.97)$      &       323     &       $4.63_{-0.46}^{+0.71}$  &       $86_{-27}^{+56} \,(2.57)$       &       20      &       $9.72_{-0.43}^{+0.67}$ &        $159_{-51}^{+72} \, (2.26)$      \\
0.20    &       2467    &       $0.986_{-0.061}^{+0.075}$       &       $17.2_{-6.2}^{+14.6} \, (2.28)$      &       275     &       $4.60_{-0.45}^{+0.82}$  &       $104_{-32}^{+75} \,(2.94)$       &       21      &       $9.29_{-0.14}^{+0.83}$ &        $220_{-67}^{+151} \,(3.10)
$       \\
0.30    &       2110    &       $0.989_{-0.063}^{+0.069}$       &       $20.5_{-7.2}^{+16.1} \, (2.55)$      &       233     &       $4.59_{-0.46}^{+0.89}$ &        $119_{-42}^{+91} \,(3.20)$       &       17      &       $9.82_{-0.51}^{+0.73}$ &        $232_{-74}^{+60} \, (2.91)$      \\
0.40    &       2046    &       $0.986_{-0.063}^{+0.075}$       &       $23.8_{-8.0}^{+18.9} \, (2.81)$      &       151     &       $4.65_{-0.50}^{+0.82}$  &       $123_{-31}^{+100} \,(3.09)$       &               &               &                       \\
0.50    &       1730    &       $0.987_{-0.064}^{+0.072}$       &       $28.6_{-9.5}^{+25.1} \, (3.16)$      &       104     &       $4.56_{-0.41}^{+0.69}$  &       $176_{-74}^{+67} \,(4.21)$       &               &               &                       \\
0.60    &       1481    &       $0.982_{-0.057}^{+0.077}$       &       $33_{-11}^{+25} \, (3.48)$      &       87      &       $4.65_{-0.54}^{+0.95}$ &        $181_{-48}^{+77} \,(4.01)$       &               &               &                       \\
0.69    &       1272    &       $0.983_{-0.059}^{+0.078}$       &       $38_{-12}^{+25} \, (3.74)$      &       51      &       $4.71_{-0.55}^{+0.65}$  &       $190_{-38}^{+69} \,(3.95)$       &               &               &                       \\
0.79    &       1067    &       $0.982_{-0.057}^{+0.077}$       &       $43_{-13}^{+28} \, (3.96)$      &       32      &       $4.46_{-0.33}^{+0.80}$ &        $219_{-47}^{+70} \,(4.49)
$       &               &               &                       \\
0.89    &       862     &       $0.985_{-0.058}^{+0.076}$       &       $50_{-15}^{+30}  \, (4.36)$     &       22      &       $4.75_{-0.60}^{+0.61}$  &       $243_{-75}^{+122} \,(4.40)$       &               &               &                       \\
1.00    &       646     &       $0.980_{-0.059}^{+0.072}$ &     $55_{-15}^{+32}  \, (4.57)$             &       11      &       $4.48_{-0.43}^{+0.51}$  &       $270_{-42}^{+63} \,(4.87)
$       &               &               &                       \\
\hline
\end{tabular}
\vspace{20pt}

\begin{tabular}{c|ccc|ccc|ccc}
\hline
\hline
$z$ & \multicolumn{3}{c|}{$fR4$-L} & \multicolumn{3}{c|}{$fR4$-I} & \multicolumn{3}{c}{$fR4$-H} \\
\hline
    & size & $ M_{200c}$ & MAR $(\Gamma_{\rm dyn})$ & size & $M_{200c}$ & MAR $(\Gamma_{\rm dyn})$ & size & $M_{200c}$ & MAR $(\Gamma_{\rm dyn})$ \\
       &    & {\small $[10^{14}\, h^{-1}$M$_\odot]$} & {\small $[10^{3}\, h^{-1}$M$_\odot]$~yr$^{-1}$} & & {\small $[10^{14}\, h^{-1}$M$_\odot]$} & {\small $[10^{3}\, h^{-1}$M$_\odot]$~yr$^{-1}$} & & {\small $[10^{14}\, h^{-1}$M$_\odot]$} & {\small $[10^{3}\, h^{-1}$M$_\odot]$~yr$^{-1}$} \\

\hline
0.00    &       3301    &       $0.993_{-0.064}^{+0.071}$       &       $11.9_{-4.1}^{+6.4} \,(1.73)$       &       652     &       $4.70_{-0.52}^{+0.74}$  &       $83_{-33}^{+43} \,(2.56)$       &       58      &       $9.85_{-0.65}^{+0.72}$ &        $211_{-73}^{+136} \, (3.09)$      \\
0.09    &       3249    &       $0.989_{-0.064}^{+0.072}$       &       $14.6_{-5.5}^{+9.2} \,(2.03)$       &       550     &       $4.72_{-0.52}^{+0.77}$& $95_{-37}^{+50} \,(2.79)$       &       55      &       $9.85_{-0.60}^{+0.70}$  &       $178_{-57}^{+128} \, (2.49)$      \\
0.20    &       3105    &       $0.989_{-0.062}^{+0.071}$       &       $17.9_{-6.8}^{+13.0} \,(2.36)$       &       432     &       $4.70_{-0.50}^{+0.73}$  &       $106_{-33}^{+73} \,(2.94)$       &       40      &       $9.79_{-0.51}^{+0.48}$  &       $272_{-98}^{+232} \, (3.63)$      \\
0.30    &       2919    &       $0.991_{-0.064}^{+0.072}$       &       $21.9_{-8.0}^{+16.3} \,(2.73)$       &       334     &       $4.63_{-0.46}^{+0.78}$  &       $130_{-41}^{+79} \,(3.45)$       &       30      &       $9.57_{-0.40}^{+0.74}$ &        $271_{-56}^{+264} \, (3.50)$      \\
0.40    &       2717    &       $0.988_{-0.060}^{+0.073}$ &     $26.0_{-9.1}^{+22.1} \,(3.06)$       &       256     &       $4.59_{-0.44}^{+0.70}$& $143_{-41}^{+115} \,(3.62)$       &               &               &                       \\
0.50    &       2351    &       $0.986_{-0.061}^{+0.076}$       &       $29.4_{-9.6}^{+22.5} \,(3.26)$       &       203     &       $4.61_{-0.47}^{+0.84}$  &       $170_{-53}^{+128} \,(4.04)$       &               &               &                       \\
0.60    &       2115    &       $0.986_{-0.060}^{+0.075}$       &       $35_{-11}^{+30} \,(3.71)$       &       135     &       $4.63_{-0.39}^{+0.64}$  &       $198_{-62}^{+105} \,(4.42)$       &               &               &                       \\
0.69    &       1742    &       $0.987_{-0.060}^{+0.078}$       &       $42_{-14}^{+32} \,(4.17)$       &       95      &       $4.58_{-0.41}^{+0.85}$& $234_{-74}^{+93} \,(5.00)$       &               &               &                       \\
0.79    &       1534    &       $0.984_{-0.061}^{+0.074}$       &       $48_{-16}^{+34} \,(4.50)$       &       57      &       $4.47_{-0.35}^{+0.89}$  &       $234_{-54}^{+94} \,(4.77)
$       &               &               &                       \\
0.89    &       1256    &       $0.988_{-0.062}^{+0.072}$       &       $54_{-16}^{+39} \,(4.72)$       &       34      &       $4.55_{-0.37}^{+0.94}$  &       $246_{-57}^{+86} \,(4.65)$       &               &               &                       \\
1.00    &       1039    &       $0.988_{-0.062}^{+0.069}$       &       $61_{-18}^{+37} \,(5.01)$       &       23      &       $4.49_{-0.29}^{+0.70}$  &       $304_{-89}^{+102} \,(5.47)$       &               &               &                       \\
\hline
\end{tabular}
\end{center}
\footnotesize{{Notes.} The upper (lower) table is for $\Lambda$CDM ($fR4$).}
\end{table*} 

\section{Results} \label{numRes}

We now describe our results. In Sect. \ref{numRes2}, we show the relation between the splashback radius $R_{\rm spl}$ and the radial location $\varsigma$ of the minimum of the derivative of the cumulative radial mass profile. In Sect. \ref{numRes3} we investigate the dependence of $\varsigma$ on redshift, MAR, and the theory of gravity.

\subsection{ Splashback radius and the radial location $\varsigma$ of the minimum of the differential mass profile} \label{numRes2}

In this section, we compute the radial logarithmic derivative of the density profiles of the synthetic clusters listed in Table \ref{samples} and identify $R_{\rm spl}$. We then compare $R_{\rm spl}$ with the radial location $\varsigma$ of the minimum of the derivative of the cumulative radial mass profiles.

For each redshift and mass bin, we computed the average radial shell mass density profile, $\rho_{\rm shell}(r)$, of the halos as follows. For each halo, we computed the mass within 200 adjacent spherical shells progressively from the center. The shells had outer cluster-centric radii spanning the range $(0.1,10) R_{200c}$, and they were logarithmically spaced. We then computed the average shell mass profile of each halo sample: The value of the mass at each radial bin $r_i$ of this profile is the mean of the individual shell masses of all the halos in the sample at $r_i$. Each value of these average profiles was determined by a number of data equal to the size of the corresponding sample (columns ``size'' in Table \ref{samples}). From these average profiles $\rho_{\rm shell}(r)$, we also computed the logarithmic derivative, ${d \log\rho_{\rm shell}/d \log r}$. This procedure automatically includes the mean background matter density $\rho_{\mathrm {bkg}}(z) = 3H^2(z)\Omega_M(z)/8\pi G$, where $\Omega_M(z)=\Omega_M(1+z)^3/\left[\Omega_M(1+z)^3+\Omega_\Lambda \right]$, and $H(z)$ and $G$ are the Hubble parameter and the gravitational constant, respectively. This procedure is the same as the approach adopted for real clusters. 

\begin{figure*}[h]
\begin{center}
\includegraphics[scale=0.53]{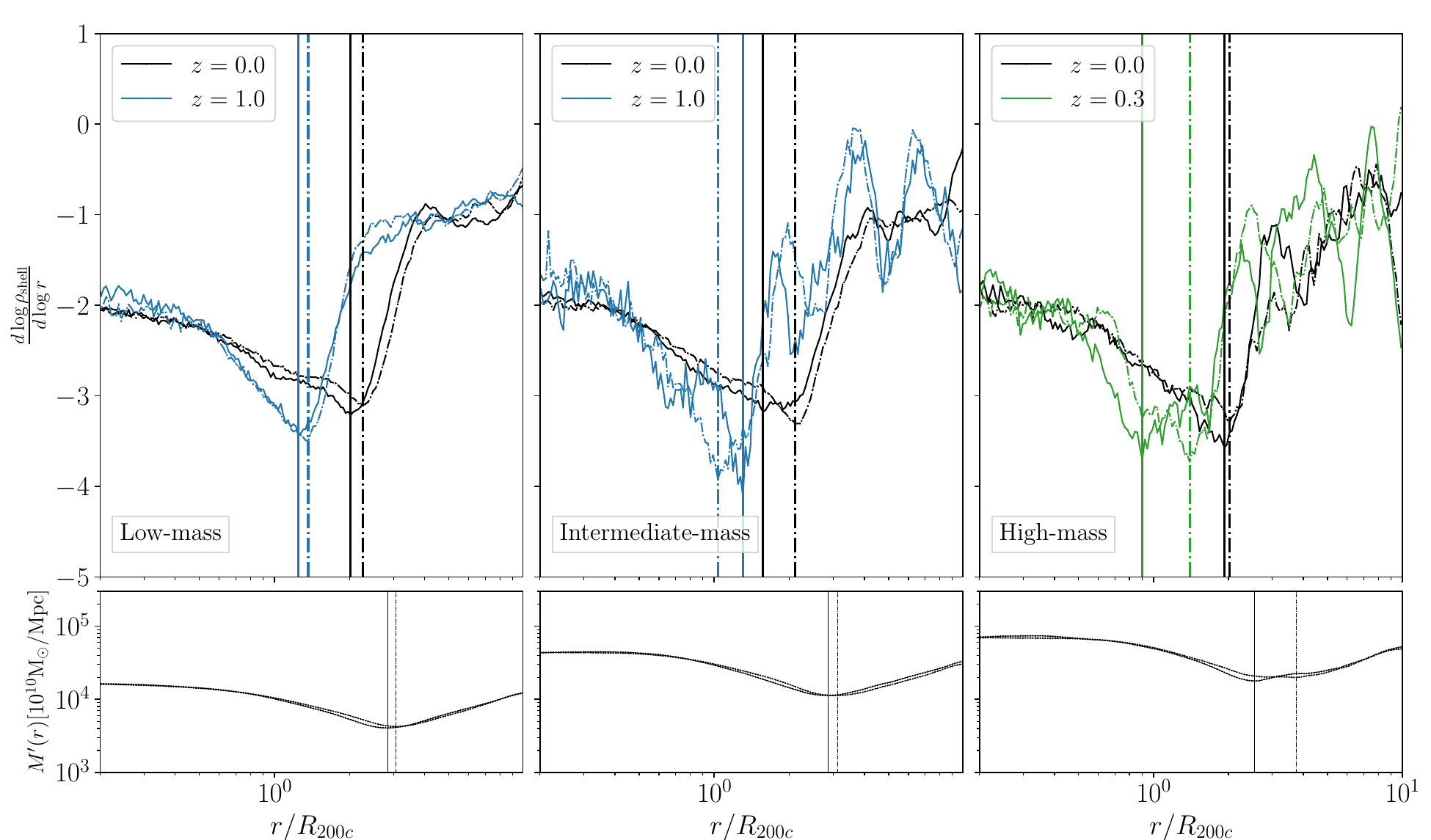} 
\caption{Logarithmic derivative of the average density profiles, and $R_{\rm spl}$ as function of mass, redshift, and theory of gravity. {Upper panels:} Average profiles of ${\mathrm {d}} \log \rho_{\rm shell}/{\mathrm{d}} \log r$ for the samples of simulated halos in Table \ref{samples}. The left, middle, and right panels show the average profiles for the low-, intermediate-, and high-mass samples, respectively, at the lowest and highest redshifts of each sample, as listed in the insets. The solid (dash-dotted) curves are the profiles in the $\Lambda$CDM ($fR4$) model.  
The vertical solid (dash-dotted) thick black lines show the location of $R_{\rm spl}$ at $z=0$ in the $\Lambda$CDM ($fR4$) model. The pair of vertical blue (green) solid and dash-dotted lines shows  $R_{\rm spl}$ at $z=1.0$ ($z=0.3$) for the $\Lambda$CDM and $fR4$ models, respectively. 
Bottom panels: The thin solid (dash-dotted) black curves show the radial derivative of the average mass profile in $\Lambda$CDM (fR4) at $z=0$; the vertical thin solid (dash-dotted) black lines show $\varsigma_{\rm avg}$ of the low-, intermediate-, and high-mass samples at $z=0$ in the $\Lambda$CDM ($fR4$) models (Table \ref{splash_samples}, first row, values in brackets). When extended to the upper panels, these thin solid and dash-dotted black lines cross the average density profiles at ${\mathrm {d}} \log \rho_{\rm shell}/{\mathrm{d}} \log r \simeq -2$, as expected from Eq. (\ref{cond2}).}\label{splashSim}
\end{center}
\end{figure*}

In Fig. \ref{splashSim}, the curves in the upper left, middle, and right panels show examples of the logarithmic derivative of the average shell density profile in the low-, intermediate-, and high-mass bins, respectively. Each profile is smoothed with a Savitzky-Golay filter \citep{SavGol64} with a window length of 25 bins. Colors and line styles show the dependence of the profiles on redshift and theory of gravity, respectively. 

All the simulated average profiles of ${d \log \rho_{\rm shell}/d \log r}$ show an absolute minimum. \citet{Adhikari2014} associated this feature with $R_{\rm spl}$, which roughly corresponds to the first apocenter of the orbits of recently accreted material. 
We identify $R_{\rm spl}$ of each halo sample as the point of absolute minimum of the logarithmic derivative of the associated smoothed average density profile $\rho_{\rm shell}(r)$. 
In the upper panels of Fig. \ref{splashSim}, each vertical line shows the point of minimum, $R_{\rm spl}$, of the logarithmic derivatives of $\rho_{\rm shell}(r)$ for the corresponding redshift (color) and theory of gravity (style).

As shown in Table \ref{splash_samples}, $R_{\rm spl}$ decreases with increasing redshift at fixed mass, and with increasing mass at fixed redshift. 
At $z\leq 0.7$, at fixed redshift and mass, $\Lambda$CDM systematically predicts a smaller $R_{\rm spl}$ than $fR4$: From $z=0$ to $z=0.7$, the relative differences between $R_{\rm spl}$ in $fR4$ and $\Lambda$CDM is in the range $\sim 2-15\%$ in the low-mass bin, and $\sim 2-35\%$ in the intermediate-mass bin. In the high-mass bin, the same difference increases from $\sim 5\%$ at $z=0$ to $\sim 55\%$ at $z=0.3$. At $z\geq 0.8$, the low-mass bin confirms this behavior: $R_{\rm spl}$ in $fR4$ is larger than in $\Lambda$CDM by $\sim 2-12\%$. In contrast, in the intermediate-mass bin, at $z=0.8-1.0$, $R_{\rm spl}$ in $\Lambda$CDM can be $\sim 22\%$ larger than in $fR4$. 

\begin{table*}[h]
\begin{center}
\caption{ \label{splash_samples} Radii $R_{\rm spl}$ and $\varsigma_{\rm avg}$ of the simulated clusters.}
\begin{tabular}{c|ccc|ccc|ccc}
\hline
\hline
$z$ & \multicolumn{3}{c|}{Low-mass samples} & \multicolumn{3}{c|}{Intermediate-mass samples} & \multicolumn{3}{c}{High-mass samples} \\
\hline
    & $\frac{{R_{\rm spl}}_{\Lambda{\text{CDM}}}}{R_{200c}}$ & $\frac{{R_{\rm spl}}_{fR4}}{R_{200c}}$ & $\frac{{R_{\rm spl}}_{fR4}-{R_{\rm spl}}_{\Lambda{\text{CDM}}}}{{R_{\rm spl}}_{\Lambda{\text{CDM}}}}$ & $\frac{{R_{\rm spl}}_{\Lambda{\text{CDM}}}}{R_{200c}}$ & $\frac{{R_{\rm spl}}_{fR4}}{R_{200c}}$ & $\frac{{R_{\rm spl}}_{fR4}-{R_{\rm spl}}_{\Lambda{\text{CDM}}}}{{R_{\rm spl}}_{\Lambda{\text{CDM}}}}$ & $\frac{{R_{\rm spl}}_{\Lambda{\text{CDM}}}}{R_{200c}}$ & $\frac{{R_{\rm spl}}_{fR4}}{R_{200c}}$ & $\frac{{R_{\rm spl}}_{fR4}-{R_{\rm spl}}_{\Lambda{\text{CDM}}}}{{R_{\rm spl}}_{\Lambda{\text{CDM}}}}$  \\

\hline
0.00    &       $2.02\,(2.85)$ & $2.26\,(3.06)$ & $0.12\,(0.07)$ &      $1.57\,(2.85)$ & $2.11\,(3.13)$ & $0.35\,(0.10)$ & $1.93\,(2.54)$ & $2.02\,(3.76)$ & $0.05\,(0.48)$    \\
0.09    &       $1.84\,(2.60)$ & $2.07\,(2.85)$ & $0.12\,(0.10)$ &      $1.93\,(2.60)$ & $1.97\,(2.79)$ & $0.02\,(0.07)$ & $1.76\,(2.21)$ & $1.97\,(2.66)$ & $0.12\,(0.20)$ \\
0.20    & $1.76\,(2.54)$ & $2.02\,(2.72)$ & $0.15\,(0.07)$ & $1.68\,(2.37)$ & $1.84\,(2.72)$ & $0.10\,(0.15)$ & $1.22\,(2.07)$ & $1.72\,(2.21)$ & $0.41\,(0.07)$ \\
0.30 & $1.72\,(2.32)$ & $1.88\,(2.54)$ & $0.10\,(0.10)$ & $1.40\,(2.11)$ & $1.72\,(2.48)$ & $0.23\,(0.17)$ & $0.90\,(1.88)$ & $1.40\,(1.93)$ & $0.55\,(0.02)$ \\
0.40    &       $1.60\,(2.26)$ & $1.76\,(2.48)$ & $0.10\,(0.10)$ & $1.46\,(2.02)$ & $1.53\,(2.21)$ & $0.05\,(0.10)$ &             &               &                       \\
0.50    &       $1.53\,(2.16)$ & $1.72\,(2.37)$ & $0.12\,(0.10)$ & $1.30\,(2.07)$ & $1.46\,(1.97)$ & $0.12\,(-0.05)$ &            &               &                       \\
0.60    &       $1.43\,(2.07)$ & $1.64\,(2.26)$ & $0.15\,(0.10)$ & $1.19\,(2.11)$ & $1.24\,(2.07)$ & $0.05\,(-0.02)$ &            &               &                       \\
0.69&   $1.46\,(2.07)$ & $1.50\,(2.11)$ & $0.02\,(0.02)$ & $1.11\,(2.16)$ & $1.36\,(1.97)$ & $0.23\,(-0.09)$ &            &               &                       \\
0.79    &       $1.33\,(1.93)$ & $1.50\,(2.07)$ & $0.12\,(0.07)$ & $1.60\,(2.02)$ & $1.24\,(2.11)$ & $-0.22\,(0.05)$ &            &               &                       \\
0.89    &       $1.36\,(1.93)$ & $1.40\,(1.93)$ & $0.02\,(0.00)$ & $1.14\,(1.76)$ & $1.33\,(2.26)$ & $0.17\,(0.29)$ &             &               &                       \\
1.00    &       $1.24\,(1.88)$ & $1.36\,(1.84)$ & $0.10\,(-0.02)$ & $1.30\,(2.37)$ & $1.04\,(1.68)$ & $-0.21\,(-0.29)$ &           &               &                       \\
\hline
\end{tabular}
\end{center}
{{Notes.} The values reported in brackets refer to the point of the minimum of the radial derivative of the average mass profile, $\varsigma_{\rm avg}$.} 
\end{table*}

A major issue in the use of the ${d \log \rho_{\rm shell}/d \log r}$ profile is that when it is computed for real clusters from the estimated mass profile, the uncertainty in the $i$th bin is proportional to ${r_i}/(r_{i+1}-r_i)$, where $r_i$ and $r_{i+1}$ are the radii used to compute the numerical derivative of the shell mass density profile, $\rho_{\rm shell}(r)$. The uncertainty is thus increasingly large at increasing radius. Furthermore, the dependence of $R_{\rm spl}$ on $\rho_{\rm shell}(r)$ hinders the robust estimation of $R_{\rm spl}$ due to the scatter caused by the high sensitivity of the logarithmic derivative to rapid changes of $\rho_{\rm shell}(r)$ in two adjacent shells. We can overcome these problems by considering, instead of $\rho_{\rm shell}(r)$ and $R_{\rm spl}$, the radial differential mass profile $dM/dr=M^\prime(r)$ and the radial location, $\varsigma$, of the minimum of this profile. 

$R_{\rm spl}$ and $\varsigma$ are not equivalent: The condition for $R_{\rm spl}$ is ${d}[{d \log \rho_{\rm shell}}/{d \log r}]/dr=0$, whereas the condition for $\varsigma$ is
\begin{equation}\label{cond2}
\centering
M^{\prime\prime}\simeq \frac{\rm d}{{\rm d} r} \left(r^2\rho_{\rm shell}\right)=0 \iff \frac{d \log \rho_{\rm shell}}{d \log r}= -2.
\end{equation}  
The values in brackets in Table \ref{splash_samples} show the points of minimum $\varsigma_{\rm avg}$ computed from the radial derivative of the averaged cumulative mass profiles of the halos in each bin. The $\varsigma_{\rm avg}$ appear larger than $R_{\rm spl}$: By averaging over all the masses and redshifts, $\varsigma_{\rm avg}$ exceeds $R_{\rm spl}$ by $43\%$ and $40\%$ in the $\Lambda$CDM and fR4 model, respectively. This result agrees with the recent study by \citet{Garcia21}, who reported a ratio of the minimum of $r^2\xi_{hm}$, where $\xi_{hm}$ is the halo-matter correlation function, and $R_{\rm spl}$ of $\sim 1.3$. 
In Fig. \ref{splashSim}, the thin solid (dash-dotted) black curves in the bottom panels show the $M^\prime$ average profiles in $\Lambda$CDM (fR4) at $z=0$; the thin black vertical solid (dash-dotted) lines show the points of minimum of these curves, namely $\varsigma_{\rm avg}$ in $\Lambda$CDM (fR4) at $z=0$. 

Despite the obvious difference between $\varsigma_{\rm avg}$ and $R_{\rm spl}$, these two quantities share remarkable similarities in their dependence on redshift, mass, and gravity model. Table \ref{splash_samples} shows that $\varsigma_{\rm avg}$, similarly to $R_{\rm spl}$, decreases with increasing redshift at fixed mass, and with mass at fixed redshift. Moreover, analogously to $R_{\rm spl}$, in all the low-mass samples the $\varsigma_{\rm avg}$ in $fR4$ are larger than in $\Lambda$CDM by $\sim 10\%$, except in the low-mass sample at $z=1.0$, where the $\varsigma_{\rm avg}$ in $fR4$ is $\sim 2\%$ smaller than in $\Lambda$CDM. Similarly, in all the high-mass samples, the $\varsigma_{\rm avg}$ in $fR4$ are larger than in $\Lambda$CDM by $5-55\%$. In the intermediate-mass bin, the similarity with $R_{\rm spl}$ also holds globally: At $z\leq 0.4$, the $\varsigma_{\rm avg}$ in $fR4$ are larger than in $\Lambda$CDM by $5-35\%$. At higher redshifts, this hierarchy can be inverted, with $\varsigma_{\rm avg}$ in $\Lambda$CDM larger up to $29\%$ than in $fR4$. 

We assessed how well $\varsigma$ can distinguish between the $\Lambda$CDM and $fR4$ models through its dependence on redshift and mass. 
To do this, in addition to $\varsigma_{\rm avg}$ extracted from the average mass profiles and discussed above, we used a different estimate of $\varsigma$  extracted from the individual mass profiles. In the simulations, the $M^\prime(r)$ profiles are less sensitive to sudden mass differences caused by random variations in the number of dark matter particles in two adjacent shells. We are thus able to exploit the $\varsigma$s of the individual halos in our analysis. 
More precisely, for each redshift and mass bin, we computed the cumulative radial mass profiles of all the halos. The profiles were obtained by progressively adding from the center the mass of 200 adjacent spherical shells with logarithmically spaced outer cluster-centric radii spanning the range $(0.1,10) R_{200c}$. 
For each halo, we computed the corresponding radial derivative of the mass profile. 

\begin{figure*}
\begin{center}
\includegraphics[scale=0.85]{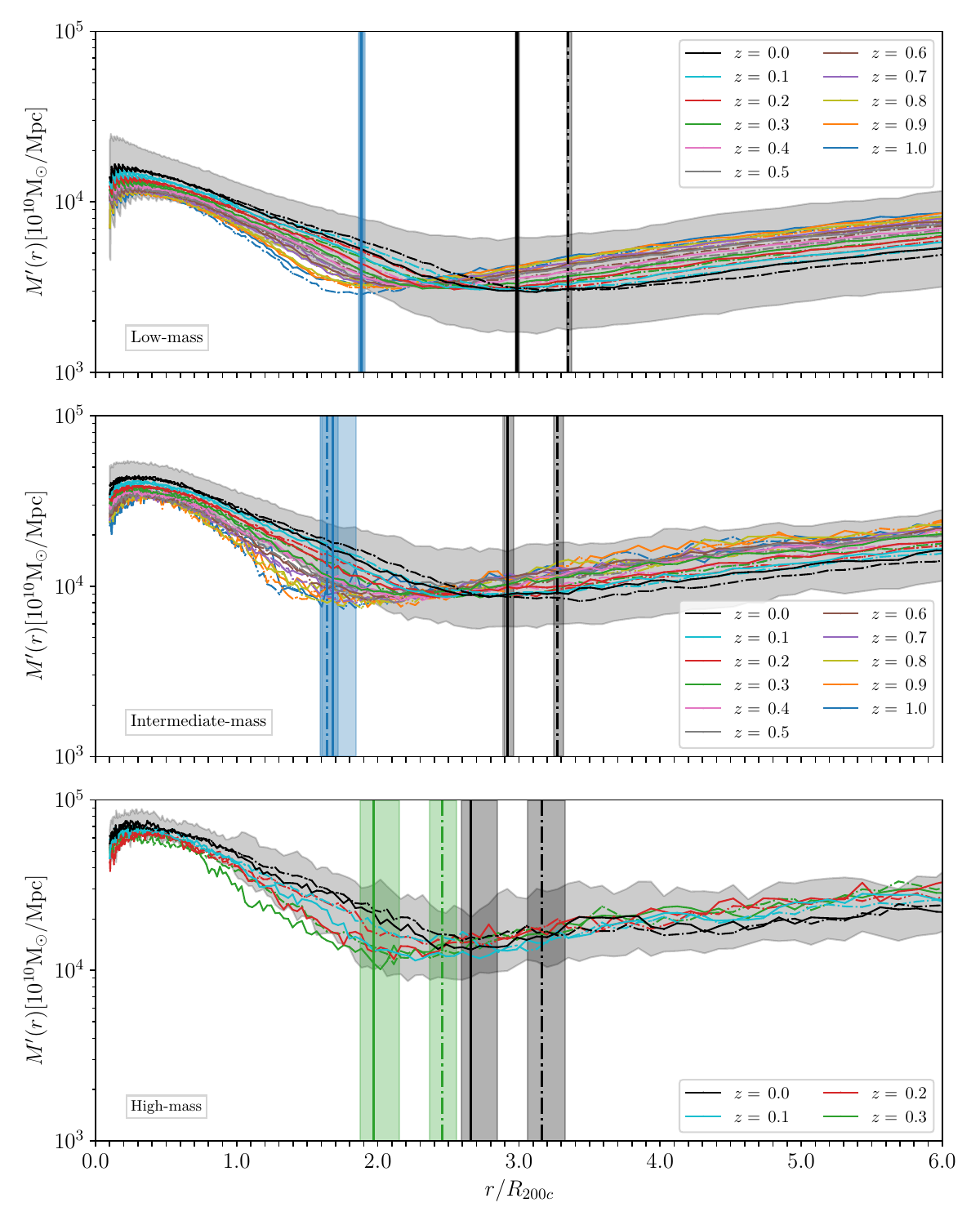} 
\caption{Median differential radial mass profiles of the samples of the simulated halos in Table \ref{samples}. The top, middle, and bottom panels show the results for the low-, intermediate-, and high-mass samples, respectively. The solid (dash-dotted) curves are the median profiles for the $\Lambda$CDM ($fR4$) halos. The profiles of the halos at different redshifts are shown with different colors, as listed in the insets. The shaded gray areas show the $68$th percentile range of the distributions of the profiles of the individual halos at $z=0$ in the $\Lambda$CDM model. The spreads of the profiles at the other redshifts and in the $fR4$ model are comparable. The vertical thick solid (dash-dotted) black line shows the median radius $\varsigma$ at $z=0$ in the $\Lambda$CDM ($fR4$) model. The vertical thick blue (gray) lines show $\varsigma$ for the profiles at $z=1.0$ ($z=0.3$): the solid and dash-dotted lines show the $\Lambda$CDM and $fR4$ models, respectively. The dash-dotted blue line in the upper panel is overplotted on the solid blue line. The vertical stripes indicate the error bands of the median $\varsigma$s.}\label{derSim}
\end{center}
\end{figure*}

In Fig. \ref{derSim}, the upper, middle, and bottom panels show the median radial differential mass profiles $M^\prime(r)$ of the low-, intermediate-, and high-mass samples of Table \ref{samples}, respectively, with their $68$th percentile ranges. The solid (dash-dotted) curves are for the $\Lambda$CDM ($fR4$) samples: for each sample, the value of the corresponding curve at each radial bin $r_i$ is the median of the set of the individual $M^\prime(r_i)$ of all the halos in the sample; each point of these curves is determined by a number of data equal to the size of the corresponding sample (columns ``size'' in Table \ref{samples}). 

\begin{figure}
    \centering
    \includegraphics[scale=0.59]{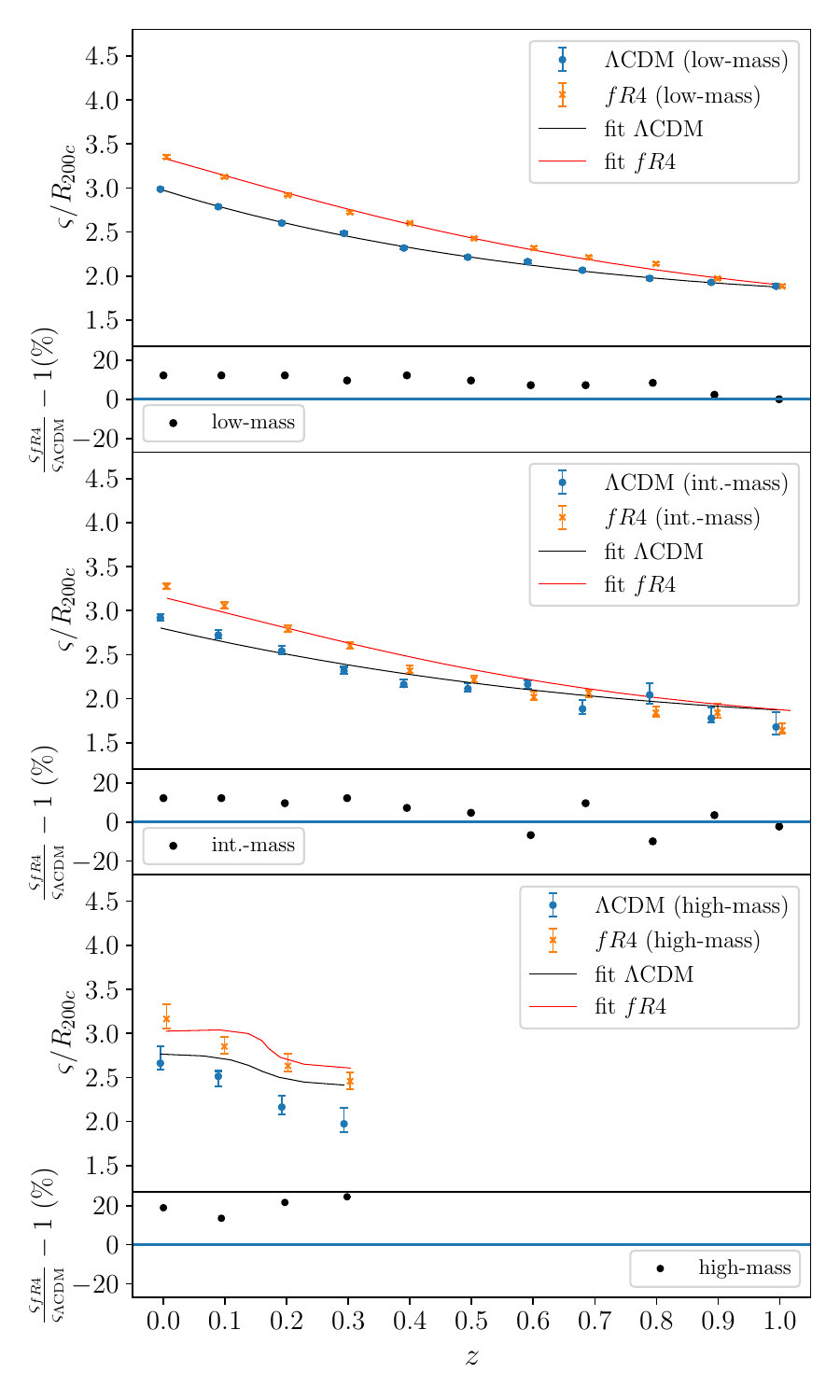} 
    \caption{Medians and associated errors of $\varsigma$ as a function of redshift in $\Lambda$CDM (blue points) and $fR4$ (orange points) for the low- (top panel), intermediate- (middle panel), and high-mass (bottom panel) samples. We applied small redshift offsets to the $fR4$ points to facilitate comparison.
    The black (red) lines are the fits of Eq. (\ref{fitRel}) to the $\Lambda$CDM ($fR4$) points: specifically, the curves are the result of a spline interpolation of the values of $\varsigma$ (11 for the low- and intermediate-mass bin, 4 for the high-mass bin) resulting from Eq. (\ref{fitRel}) using the redshift, mass, and mass accretion rate of each simulated bin (Table \ref{samples}).
    In each panel, the lower plot shows the differences of the median $\varsigma$s in $\Lambda$CDM and $fR4$ relative to the $\Lambda$CDM value.}
    \label{fig:compMin}
\end{figure}

We identified the final $\varsigma$ of each mass and redshift sample as the median of the set of all the $\varsigma$s of the individual halos in the sample. The $\varsigma$ of each individual halo is the minimum of the corresponding profile $M^\prime(r)$ beyond $0.5 R_{200c}$. This choice avoids possible inconsistencies caused by high fluctuations in the inner region of the halo. In Fig. \ref{derSim}, the solid (dash-dotted) vertical lines show the value of $\varsigma$ in $\Lambda$CDM ($fR4$) at the lowest and highest redshift of the three mass bins. In general, the position of $\varsigma$ does not coincide with the minimum of the corresponding median profile in Fig. \ref{derSim} because we did not select $\varsigma$ at the minimum of the median profile of the individual halos, but as the median of the $\varsigma$s of the individual halos.
Table \ref{ptmin_samples} lists the median $\varsigma$ of the individual differential mass profiles for the samples listed in Table \ref{samples}. We also provide an estimate of the error of the median with the difference between the median and the corresponding 16th percentile, and the difference between the median and the 84th percentile value divided by the square root of the sample size. We also display the relative differences of $\varsigma$ in $\Lambda$CDM and $fR4$. 
Figure \ref{fig:compMin} shows $\varsigma$ as a function of redshift and mass in $\Lambda$CDM and $fR4$.
    
\begin{table*}[h]
\begin{center}
\caption{ \label{ptmin_samples} { $\varsigma$ of the simulated clusters.}}
\begin{tabular}{c|ccc|ccc|ccc}
\hline
\hline
$z$ & \multicolumn{3}{c|}{Low-mass samples} & \multicolumn{3}{c|}{Intermediate-mass samples} & \multicolumn{3}{c}{High-mass samples} \\
\hline
    & $\frac{\varsigma_{\Lambda{\text{CDM}}}}{R_{200c}}$ & $\frac{\varsigma_{fR4}}{R_{200c}}$ & $\frac{\varsigma_{fR4}-\varsigma_{\Lambda{\text{CDM}}}}{\varsigma_{\Lambda{\text{CDM}}}}$ & $\frac{\varsigma_{\Lambda{\text{CDM}}}}{R_{200c}}$ & $\frac{\varsigma_{fR4}}{R_{200c}}$ & $\frac{\varsigma_{fR4}-\varsigma_{\Lambda{\text{CDM}}}}{\varsigma_{\Lambda{\text{CDM}}}}$ & $\frac{\varsigma_{\Lambda{\text{CDM}}}}{R_{200c}}$ & $\frac{\varsigma_{fR4}}{R_{200c}}$ & $\frac{\varsigma_{fR4}-\varsigma_{\Lambda{\text{CDM}}}}{\varsigma_{\Lambda{\text{CDM}}}}$  \\

\hline
0.00    &       $2.986_{-0.013}^{+0.020}$ & $3.350_{-0.013}^{+0.022}$ & 0.122 &       $2.918_{-0.027}^{+0.046}$ & $3.274_{-0.027}^{+0.046}$
& 0.122 & $2.661_{-0.070}^{+0.19}$ & $3.16_{-0.10}^{+0.16}$ & 0.189     \\
0.09    &       $2.786_{-0.012}^{+0.019}$       & $3.126_{-0.013}^{+0.019}$  & 0.122  &     $2.723_{-0.034}^{+0.053}$ & $3.055_{-0.031}^{+0.045}$
& 0.122 & $2.512_{-0.11}^{+0.061}$      & $2.851_{-0.079}^{+0.11}$ & 0.135 \\
0.20    &       $2.600_{-0.012}^{+0.018}$ & $2.918_{-0.013}^{+0.018}$ & 0.122 & $2.541_{-0.037}^{+0.058}$ & $2.786_{-0.028}^{+0.043}$
& 0.096 & $2.163_{-0.087}^{+0.13}$ & $2.631_{-0.064}^{+0.139}$ & 0.216 \\
0.30&   $2.483_{-0.012}^{+0.017}$ & $2.723_{-0.012}^{+0.018}$ & 0.096 & $2.318_{-0.037}^{+0.044}$ & $2.600_{-0.032}^{+0.041}$
& 0.122 & $1.973_{-0.097}^{+0.18}$ & $2.456_{-0.089}^{+0.10}$ & 0.245 \\
0.40    &       $2.318_{-0.011}^{+0.018}$ & $2.600_{-0.012}^{+0.016}$ & 0.122 & $2.163_{-0.026}^{+0.051}$ & $2.318_{-0.030}^{+0.055}$
& 0.072 &               &               &                       \\
0.50    &       $2.213_{-0.013}^{+0.017}$ & $2.427_{-0.012}^{+0.017}$ & 0.096 & $2.114_{-0.035}^{+0.060}$ & $2.213_{-0.029}^{+0.049}$
& 0.047 &               &               &                       \\
0.60    &       $2.163_{-0.014}^{+0.016}$ & $2.318_{-0.012}^{+0.018}$ &  0.072 & $2.163_{-0.053}^{+0.041}$ & $2.019_{-0.034}^{+0.064}$
& -0.067 &              &               &                       \\
0.69&   $2.066_{-0.013}^{+0.018}$ & $2.213_{-0.012}^{+0.019}$ & 0.072 & $1.884_{-0.054}^{+0.10}$ & $2.066_{-0.047}^{+0.037}$
& 0.096 &               &               &                       \\
0.79    &       $1.973_{-0.014}^{+0.019}$ & $2.138_{-0.013}^{+0.017}$ & 0.084 & $2.043_{-0.097}^{+0.13}$ & $1.841_{-0.041}^{+0.071}$
& -0.099 &              &               &                       \\
0.89    &       $1.928_{-0.015}^{+0.021}$ & $1.973_{-0.012}^{+0.018}$ & 0.023 & $1.779_{-0.052}^{+0.12}$ & $1.841_{-0.057}^{+0.098}$
& 0.035 &               &               &                       \\
1.00    &       $1.884_{-0.018}^{+0.024}$ & $1.884_{-0.013}^{+0.017}$ & 0.000 & $1.679_{-0.089}^{+0.17}$ & $1.641_{-0.041}^{+0.079}$
& -0.023 &              &               &                       \\
\hline
\end{tabular}
\end{center}
\end{table*}

These new estimates of $\varsigma$ confirm the previous trends shown by $\varsigma_{\rm avg}$, but the results are less prone to large scatter: $\varsigma$ decreases with increasing redshift and mass; furthermore, at fixed redshift and mass, $\Lambda$CDM generally predicts a smaller $\varsigma$ than $fR4$.
As shown in Table \ref{ptmin_samples}, in the low- and high-mass bins, $\varsigma$ is systematically larger in $fR4$ than in $\Lambda$CDM: In the low-mass bin, the relative difference decreases with redshift from $\sim 12\%$ at $z=0$ to $0\%$ at $z=1$; in the high-mass bin, the $\varsigma$s in $fR4$ are increasingly larger than in $\Lambda$CDM as redshift increases, from $\sim 19\%$ at $z=0$ to $\sim 25\%$ at $z=0.3$. In the intermediate-mass bin, $\varsigma$ is systematically larger in $fR4$ at redshift $z\sim 0-0.5$, where the relative difference decreases from $\sim 12\%$ to $\sim 5\%$. In this mass bin, similarly to $R_{\rm spl}$ (Table \ref{splash_samples}), $\Lambda$CDM can present larger $\varsigma$s than $fR4$ at $z>0.5$.

\subsection{Dependence of $\varsigma$ on redshift, MAR, and the theory of gravity} \label{numRes3}

In the previous section, we showed that $\varsigma$, analogously to $R_{\rm spl}$, is sensitive to the redshift and the mass of the cluster, and that it can be up to 25\% larger in $fR4$ than in $\Lambda$CDM (Table \ref{ptmin_samples}). Mass and MAR are tightly correlated \citep{Diemer2014,pizzardo2020}. Our findings therefore agree with previous results showing that $R_{\rm spl}$ is correlated with redshift and MAR \citep{More2015}. In this section, we model the behavior of $\varsigma$ with $z$ and MAR as independent variables, and we investigate whether $\varsigma$ can distinguish between $\Lambda$CDM and $fR4$. 

Our definition of the MAR can be easily extended to real clusters because it does not rely on the merger history of the halos, as is usually adopted in investigations based on N-body simulations. 
We considered the procedure by \citet{pizzardo2020, pizzardo22}, who measured the MAR of $\sim 500$ real clusters of galaxies. The recipe is based on the assumption that the infall of new material onto the clusters is described by the spherical accretion model \citep{deBoni2016}. Namely, in the infall time $t_{\rm inf}$, a cluster accretes all the material within a spherical shell, centered on the cluster center, with inner radius $R_i$ and thickness $\Delta_s=\delta_s R_i$. The quantity $\delta_s$ is derived from the equation of motion of the infalling mass with initial velocity $v_{\rm inf}$ and constant acceleration $-GM(<R_i)/(R_i+\delta_sR_i/2)^2$,
\begin{equation}\label{eq:thickness}
t_{\rm inf}^2  GM(<R_i) - t_{\rm inf} 2 R_i^2 (1 + \delta_s/2)^2 v_{\rm inf} - R_i^3  \delta_s (1 + \delta_s/2)^2 = 0,
\end{equation}
where $M (< R_{i})$ is the mass of the cluster within 
$R_i$, and $G$ is the gravitational constant. Here, according to \citet{deBoni2016} and \citet{pizzardo2020,pizzardo22}, we fixed $t_{\rm inf}=1$ Gyr and $R_{i}=2 R_{200c}$, whereas for the infall velocity, $v_{\rm inf}$, we relied on the radial velocity profiles of the simulated halos of DUSTGRAIN in the two theories of gravity. For a sample of simulated clusters, $v_{\rm inf}$ is the minimum velocity of the median radial velocity profile in the range $[2,2.5] R_{200c}$ of the sample, whereas for a sample of real clusters, $v_{\rm inf}$ results from suitable interpolations of the simulated samples \citep[see][for further details]{pizzardo2020,pizzardo22}.
When the mass of the falling shell, $M_{\rm shell}$, is known, the MAR is estimated with the equation
\begin{equation}\label{mar}
{\rm MAR} \equiv \frac{M_{\rm shell}}{t_{\rm inf}}. 
\end{equation}

\begin{figure}
\begin{center}
\includegraphics[scale=0.58]{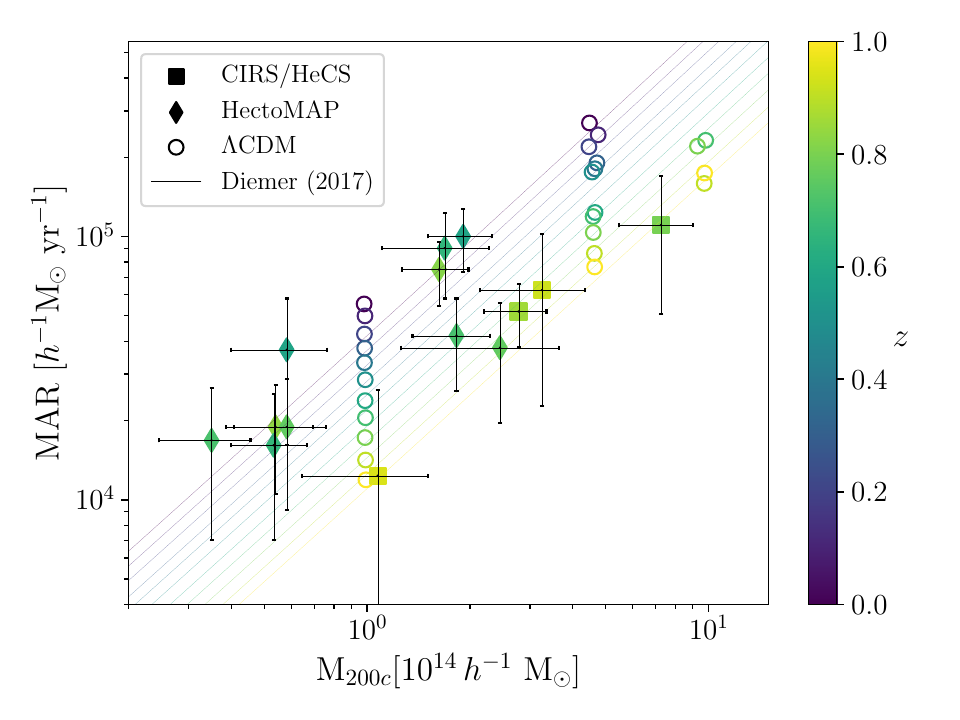}
\caption{MAR as a function of the cluster mass $M_{200c}$ in the simulations and the MAR of the CIRS, HeCS, and HectoMAP clusters. The open circles show the median MARs of the 26 $\Lambda$CDM samples of Table \ref{samples}, color-coded by redshift. The colored curves are the relation of \citet{Diemer2017sparta2} computed at the 11 redshifts of our simulated samples. The squares and the diamonds with error bars show the MAR estimates with their uncertainties of the stacked clusters of the CIRS and HeCS \citep{pizzardo2020} and HectoMAP clusters \citep{pizzardo22}, respectively.}\label{mar-compa}
\end{center}
\end{figure}

We computed the MARs of all the halos in the samples of Table \ref{samples}. This table lists the median MAR of each sample with their corresponding $68$th percentile range.
In Fig. \ref{mar-compa}, the open circles show the median MARs of the 26 $\Lambda$CDM samples as a function of the median mass of the sample, color-coded by redshift. By considering the merging history of simulated CDM halos, \citet{Diemer2017sparta2} proposed an accretion law that we computed at our 11 simulated redshifts and show with the curves in Fig. \ref{mar-compa}. The agreement between this law and our estimates shows that the MARs of galaxy clusters estimated with our recipe, which can be applied to real clusters, are consistent with the MARs derived from the merger trees, which remain inaccessible to observations. The estimated MARs of the CIRS and HeCS clusters \citep[][squares in Fig. \ref{mar-compa}]{pizzardo2020} and of the HectoMAP clusters \citep[][diamonds in Fig. \ref{mar-compa}]{pizzardo22} agree with the theoretical expectations of $\Lambda$CDM within the uncertainties. Both $M_{200c}$ and MAR are estimated from the same caustic mass profile, hence their errors are not independent. However, the error of the caustic mass profile at a given radius is determined by the number and the distribution of the galaxies within the spherical shell centered at this radius, and $M_{200c}$ and MAR are estimated at well-separated radii. Therefore, we can assume that the errors on these quantities are independent, and we obtain a roughly correct estimate of the expected uncertainty ranges (error bars in Fig.~\ref{mar-compa}). 

For each of the two sets of simulations, we fit the 26 points in the four-dimensional space $(\varsigma,z,\text{MAR}, M_{200c})$ with the relation
\begin{equation}\label{fitRel}
\frac{\varsigma}{R_{200c}}(z, \text{MAR}, M_{200c}) = a \left[1 + b\, \Omega_M(z)\right] \left[1 + c\, \exp\left(-\frac{\Gamma_{\rm dyn}}{d}\right) \right], 
\end{equation}  
where $\Omega_M(z)=\Omega_M (1+z)^3/[\Omega_M (1+z)^3 + \Omega_\Lambda]$, $\Gamma_{\rm dyn}$ is the dynamical MAR $\Gamma_{\rm dyn}\equiv {{\rm d}\log M_{\rm vir}}/{{\rm d}\log \hat a}$, with $\hat a=(1+z)^{-1}$ the scale factor, and  $a,b,c$, and $d$ are four free parameters. Figure \ref{mar-compa} proves that the difference between the $M_{\rm vir}$s at two different epochs is comparable with the difference between the masses of the infalling shell that we adopted in our definition of MAR. With an obvious change in the variables, we can therefore express the dynamical accretion rate as $\Gamma_{\rm dyn}\approx [\mathcal{K}(t, \Omega_M, H_0)/M_{200c}]\, ({\rm d} M/{\rm d}t)$, with ${{\rm d} M}/{{\rm d}t}\equiv\text{MAR}.$\footnote{ $\mathcal{K}$ transforms the time coordinates from the scale factor to cosmic time: $d \ln \hat a = \mathcal{K}^{-1}(t, \Omega_M, H_0) dt$.} Table \ref{samples} reports $\Gamma_{\rm dyn}$ for each cluster sample. Equation (\ref{fitRel}) was proposed by \citet{More2015} to model $R_{\rm spl}$ rather than $\varsigma$. The similar behavior of these two quantities with redshift, MAR, and mass leads us to heuristically use the same functional form for $\varsigma$, however.
We also kept $\Omega_M(z)$ in the form of $\Lambda$CDM in the fit to the $fR4$ data in order to direct any difference to the four fitting parameters $a,b,c$, and $d$.
We fit the median points with a nonlinear least-squares Marquardt-Levenberg algorithm by weighting the points with the inverse of their variance (Table \ref{ptmin_samples}). The values of the coefficients are reported in Table \ref{fit}.
\begin{table}[!h]
\begin{center}
\caption{ \label{fit} {Best-fit parameters of Eq. (\ref{fitRel}).}}
\begin{tabular}{ccc}
\hline
\hline
parameter & $\Lambda$CDM & $fR4$ \\
&  & \\
\hline
$a$ & $3.08\pm 0.27$ & $1.9\pm 1.8$ \\
$b$ & $-0.507\pm 0.057$ & $-0.518\pm 0.059$\\
$c$ & $0.80\pm 0.36$ & $1.5\pm 2.2$ \\
$d$ & $1.03\pm 0.49$ & $7.2\pm 7.0$ \\
\end{tabular}
\end{center}
\end{table} 

Table \ref{fit} shows that the $\Lambda$CDM fit is robust. The parameter $b$, which controls the dependence of $\varsigma$ on redshift, is constrained with a relative uncertainty of $\sim 10\%$; the parameters $c$ and $d$, which control the dependence of $\varsigma$ on the logarithmic MAR, are constrained with uncertainties of $\sim 45-47\%$. This result shows that in $\Lambda$CDM the dependence of $\varsigma$ on redshift and MAR are statistically significant and are similar to those expected for $R_{\rm spl}$. The $fR4$ fit is less strongly constrained: Except for the parameter $b$ with a $\sim 10\%$ uncertainty, the remaining three parameters are affected by uncertainties of $\sim 100\%$ or larger. However, the model of Eq. (\ref{fitRel}) was only proposed to fit the relation in the $\Lambda$CDM universe. The large uncertainties on the parameters for the $fR4$ fit can thus simply suggest that Eq. (\ref{fitRel}) is less adequate for this gravity model.

\begin{figure}
    \centering
    \includegraphics[scale=0.59]{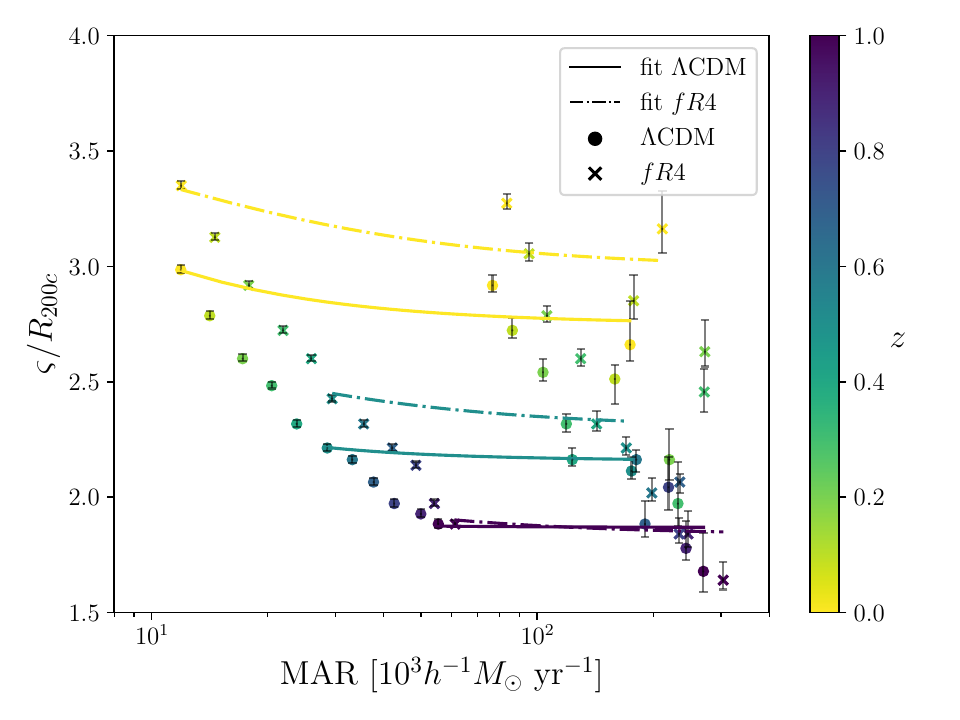}
    \caption{$\varsigma$ as a function of MAR, color-coded by redshift. The circles and crosses show the median values of the $\Lambda$CDM and $fR4$ samples, respectively. The solid (dash-dotted) curves are the fits of Eq. (\ref{fitRel}) to the $\Lambda$CDM ($fR4$) data at three redshifts. From top to bottom, $z=0.0,0.5$, and $ 1.0$. 
    At fixed redshift, $\varsigma$ decreases with increasing MAR; at fixed MAR, $\varsigma$ decreases as the redshift increases.}
    \label{fig:corr}
\end{figure}

The absolute value of the relative differences between the fit and the actual values is $\sim 4.2\%$ and $\sim 3.5 \%$ for the $\Lambda$CDM and the $fR4$ samples on average, respectively. Figure \ref{fig:compMin} shows examples of these fits in the $\varsigma-z$ plane. Figure \ref{fig:corr} shows projections of the fits on the $\varsigma-$MAR plane at three different redshifts. 
The correlation of $\varsigma$ with redshift and MAR is confirmed by the Kendall test, which returns a tight correlation with a $p$-value $10^{-5}$  for the relation $\varsigma - z$ and for the relation $\varsigma - $MAR.

As expected, the fits show that the difference between the $\Lambda$CDM and $fR4$ models is larger at low redshifts, independently of the MAR. The difference steadily decreases with increasing redshift: consequently, because at low redshift $\varsigma$ is higher in $fR4$ than in $\Lambda$CDM, the fits in the  $(\varsigma, z)$ plane at fixed MAR (Fig. \ref{fig:compMin}) show steeper profiles in $fR4$ than in $\Lambda$CDM.

\begin{figure}
\begin{center}
\includegraphics[scale=0.6]{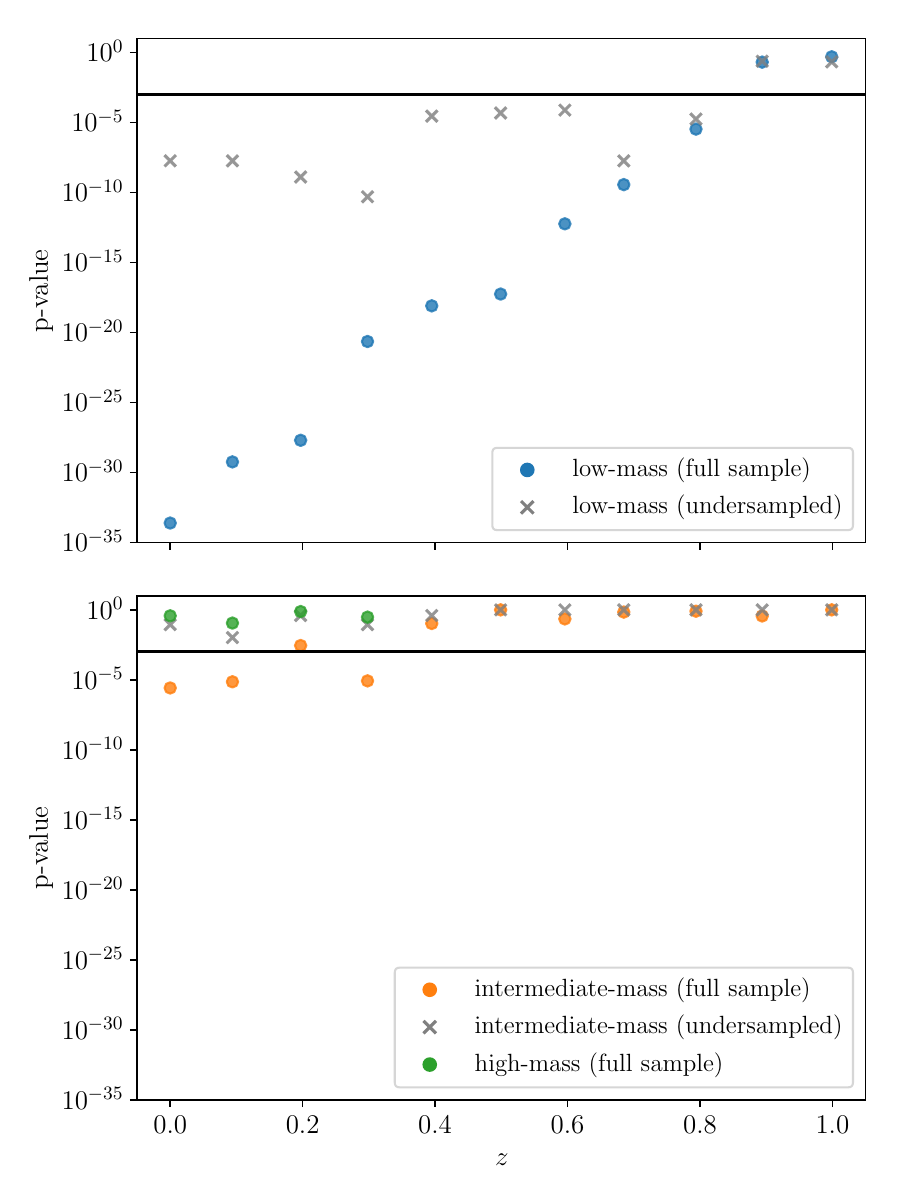} 
\caption{$p$-values of the median test to distinguish the two theories of gravity performed on the samples of $\varsigma$s of the simulated clusters in Table \ref{samples}. {\it Top panel}: Blue points show the $p$-values of the test on the full samples of $\varsigma$s in the low-mass bin, and the gray crosses show the $p$-values of the test performed by undersampling all the low-mass samples to 646 halos. {\it Bottom panel}: Orange and green points show the $p$-values of the test on the full samples in the intermediate- and high-mass bin, respectively; the gray crosses show the $p$-values of the test performed by undersampling the intermediate-mass samples to 11 halos. In both panels, the horizontal lines denote $p=10^{-3}$.}\label{med_test}
\end{center}
\end{figure}

We now quantitatively assess to which extent $\varsigma$ can distinguish between $\Lambda$CDM and $fR4$. We applied the median test to each of our 26 $\Lambda$CDM$-fR4$ pairs of the distributions of the individual $\varsigma$ at equal redshift and mass bin. This test counts the number of elements of the two distributions that are either larger or smaller than the grand median of the two samples. Based on a Pearson $\chi$-squared test, the test assigns a $p$-value to the null-hypothesis that the samples are drawn from the same distribution. Generally, the $fR4$ sample is more densely populated than the corresponding sample in $\Lambda$CDM (see Table \ref{samples}). To perform the test on samples with the same number of halos, we therefore randomly undersampled each $fR4$ sample to the number of halos in the corresponding $\Lambda$CDM sample. The results of the median test are shown in Fig. \ref{med_test}: the blue points in the upper panel are for the low-mass samples, whereas the orange and green points in the lower panel are for the intermediate- and high-mass samples, respectively.

We took $p=10^{-3}$ as the upper limit to conclude that the two theories of gravity can be distinguished: Values below this limit are evidence at a significance level higher than $3\sigma$  that the two samples are drawn from different populations. Figure \ref{med_test} shows that the low-mass samples can generally distinguish between the two models in the redshift range $z\sim 0-0.8$. The samples at the lowest redshifts show extremely strong evidence that gravity is modified $(p\approx 10^{-30})$. In contrast, only 3 of 11 intermediate-mass samples show some evidence of modified gravity, and none of the high-mass samples is able to do so. 

The progressively weaker discerning power of the median test in the low-mass samples as  redshift increases is due to the intrinsic degeneracy of $fR4$ and $\Lambda$CDM at high redshift. We illustrate in Appendix \ref{append} that the main differences between the two models are more evident at late times. This conclusion is also supported by the steady decrease in the relative difference between the $\varsigma$s of the two models as redshift increases (see Table \ref{ptmin_samples}). 
In contrast, the largest part of the intermediate- and high-mass samples shows $p>10^{-3}$ even at low redshift, where the relative difference between the values of $\varsigma$ is similar to or larger than for the low-mass halos (Table \ref{ptmin_samples}). The reason for this behavior is the small number of halos in these mass samples and not the intrinsic similarity of the models.

To test the effect of poor sampling, we considered for the low- and intermediate-mass bins the corresponding sample with the lowest number of halos, regardless of redshift and model: Table \ref{samples} shows that the smallest samples are the $\Lambda$CDM samples at $z=1.0$, with 646 and 11 halos for the low- and intermediate-mass bins, respectively. We therefore performed a median test in which we randomly undersampled all the samples of the low- and intermediate-mass bins to 646 and 11 halos, respectively. 

The results are shown by the gray crosses in Fig. \ref{med_test}.
In the low-mass bin, the undersampling increases the $p$-value of the median test for the samples at $z\sim 0-0.8$; furthermore, the increment of the $p$-value decreases with redshift because the higher the redshift, the fewer halos are removed. Nevertheless, at $z< 0.9$, although they are substantially reduced, the $p$-values remain below $10^{-3}$. 

The important result of this test is the minimum size of the sample required to distinguish between the two theories of gravity: As long as the sample size is $\gtrsim 400$, the distinguishing power remains substantially unaffected. 
Therefore, in the low-mass bin, the loss of significance of the median test to distinguish the theory of gravity at high redshifts, $z\sim 0.9-1.0$, is caused by the intrinsic degeneracy between $fR4$ and $\Lambda$CDM at these early times and not by the sample size.
Conversely, the failure of the distinguishing ability of $\varsigma$ in the intermediate- and high-mass bins is caused by the small halo samples. As long as the sample size is $\sim 230-400$, as in the first four redshift bins of the full intermediate-mass bin, the distinguishing power is still present. With smaller sample sizes, however, the $p$-values are aligned to the results of the intermediate-mass full samples at $z\gtrsim 0.4$, and of the high-mass full sample, where we have $\sim 10-150$ halos. 

\section{Discussion}\label{discussion}

We have investigated the radial location, $\varsigma$, of the minimum of the radial differential mass profile, $M^\prime(r)$, of simulated galaxy clusters halos. Although not equivalent, we show that $\varsigma$ shares similarities with the splashback radius, $R_{\rm spl}$. We show that $\varsigma$ can distinguish among different gravity models. We investigated General Relativity, with the standard $\Lambda$CDM model, and $f(R)$ gravity.
Our work strengthens the hypothesis that galaxy clusters might be used as cosmological probes. In particular, we showed that the outskirts of clusters promise to effectively complement more traditional cluster-based cosmological tests, such as cluster counts and the gas fraction \citep[see][for a review of the tests of gravity with galaxy clusters]{cataneo2020tests}. 

Like $R_{\rm spl}$, $\varsigma$ can be located directly from the mass profiles of galaxy clusters. Most works that searched for $R_{\rm spl}$ relied on the projected galaxy number density profile or on the projected luminosity profile. These quantities are tightly related and are good proxies of the three-dimensional matter density profile in general \citep{carlberg1997average,Rines2003Acairns,Rines04,Rines2006CIRS,Rines2013HeCS}. However, many studies have pointed out that the galaxy number density and luminosity profiles are not directly proportional to each other \citep[e.g.][]{Proctor15}. In addition, their correlation is sensitive to the cluster mass and to the selection function \citep{Mulroy14}. Moreover, if the galaxy density is constrained by photometrically detected clusters, projection effects and member mismatch become increasingly relevant in the outer regions \citep[e.g.,][]{More16,Baxter17,Shin19}.
The use of the mass profile to detect $R_{\rm spl}$ or $\varsigma$ is clearly preferred because the definitions of these quantities are based on the mass profile that is measured in N-body simulations. 

Furthermore, the detection of $\varsigma$ from mass profiles is prone to smaller fluctuations than the detection of $R_{\rm spl}$ because the radial derivative $M'(r)$ is much less noisy than the logarithmic slope of the matter density. For this reason, unlike $R_{\rm spl}$, $\varsigma$ can be located in individual clusters if the mass profile is accurately estimated up to $\sim 3-4 R_{200}$.

Using $\varsigma$ as a probe of the theory of gravity requires an unbiased estimate of hundreds of cluster mass profiles up to large cluster-centric radii with high precision and accuracy. Weak gravitational lensing \citep{bartelmann2010gravitational,hoekstra2013masses,Umetsu20essay} and the caustic technique \citep{Diaferio1997,Diaferio99,Serra2011} are the only methods that can be reliably used beyond the inner region of clusters, where dynamical equilibrium does not hold. 

The current cluster samples do not allow us to robustly locate $\varsigma$ and to statistically use it as a probe of the theory of gravity. However, future facilities such as the VRO \citep{Ivezic19} and the Euclid mission \citep{Laureijs11,Sartoris16} will provide weak-lensing mass measurements of thousands of clusters.
Powerful spectrographs such as the multi-object William Herschel Telescope Enhanced Area Velocity Explorer spectrograph on the WHT \citep[WEAVE,][]{Balcells10,Dalton12,Dalton14,Dalton16} will explore the infall regions of galaxy clusters and provide dense spectroscopic surveys out to radii $\lesssim 5 R_{200c}$ of more than 100 clusters at redshift $\lesssim 0.5$. Planned observations with the Prime Focus Spectrograph on Subaru \citep[PFS,][]{Tamura16} and the Maunakea Spectroscopic Explorer on the CFHT \citep[MSE,][]{MSE19} will provide spectroscopic redshifts of hundreds to thousands of galaxy cluster members for thousands of individual clusters at $z\lesssim 0.6$. These surveys will allow a precise and unbiased determination of $\varsigma$ from weak lensing and the caustic technique.

These future observations will enable synergies between photometric and spectroscopic techniques. These synergies will be crucial for promoting $\varsigma$ as a probe of gravity. At cluster-centric distances of about $\varsigma$, a misidentification of background galaxies and large-scale structure can severely bias weak-lensing mass profiles obtained from photometric redshifts \citep{Hoekstra2003,Hoekstra2011,Becker11,Sommer22}. The caustic technique provides unbiased mass profiles on average, but the scatter is relevant \citep{Serra2011,Pizzardo23}. The joint use of these two techniques could take advantage of the respective strengths, and it would be crucial to obtain exquisitely robust mass estimates up to large cluster-centric distances \citep[e.g.][]{dellantonio20}.

As a cautionary note about our analysis presented in this work, we recall that the $\Lambda$CDM and $fR4$ runs of the DUSTGRAIN simulations that we used here were normalized at the CMB epoch. Therefore, their power-spectrum normalization at $z=0$, $\sigma_8$ was 0.847 and 0.967 for the two models (Sect. \ref{sim}). To show that our results are not affected by these different values of $\sigma_8$, we compared the median MARs at $z=0$ of our low- and high-mass $\Lambda$CDM samples with the median MARs estimated by \citet{pizzardo2020,pizzardo22} in the $\Lambda$CDM L-CoDECS N-body simulation of \citet{Baldi2012CoDECS}, which has $\sigma_8=0.809$. Despite the lower $\sigma_8$, we find that the MARs in the L-CoDECS  simulation are $\sim14\%$ and $\sim 3\%$ larger than the MARs of our DUSTGRAIN samples for the low- and the high-mass bin, respectively. Moreover,  $\varsigma$  is  $\sim 4.5\%$ smaller in the DUSTGRAIN clusters than in the L-CoDECS simulations for both the low- and high-mass bin. In addition, in DUSTGRAIN, $\sigma_{8,\Lambda{\rm CDM}} < \sigma_{8,fR4}$, which means that if $\varsigma$ were inversely proportional to $\sigma_8$, we should expect $\varsigma$s larger in $\Lambda$CDM than in $fR4$ at fixed redshift and mass. This expectation is the opposite of the results from the simulations (Table \ref{ptmin_samples}). Hence, the behavior of $\varsigma$ does not appear to be correlated with $\sigma_8$, and therefore, the differences between the models that we find here are mainly driven by the different theory of gravity.

{One might expect that the outer regions of clusters smoothly approach the uniform cosmic background and that this background contributes to the mass profile of the clusters. However, the distribution of  matter around both simulated \citep[e.g.,][]{gao2004sub,vandenbosch2005,Cornwell23a,Cornwell23b} and real clusters \citep[e.g.,][]{Beers1983,Dressler1988,Martinez16,Sarron19} is inhomogeneous, with filaments, groups of galaxies, and low-density regions.
Therefore, we did not subtract the contribution of a uniform background from the mass profile the clusters in our analysis (see Sect. \ref{numRes2}). 
Nevertheless, we verified that the effect of subtracting a uniform background with density $\Omega_M\rho_c$ from the mass profiles has a negligible effect on our results: The radii $\varsigma$ at which the minimum of the radial derivative of the cluster mass profile occurs are $\sim 10\%$ larger on average than the $\varsigma$s from the profiles without subtraction. More importantly, the correlations between $\varsigma$, redshift, and MAR are preserved at similar significance levels.}

We showed that extended ranges of mass and redshift are critical to avoid possible degeneracies between the models. Thus, our analysis pointed out the importance of further efforts toward the realization of simulations of large cosmological volumes in modified gravity.
{Larger samples that include the most massive clusters can help us to establish the correlation between $\varsigma$ and MAR with more significance because these samples will allow us to stack clusters based on mass and to probe a more extended range of MARs at fixed redshift. Furthermore,} the most massive clusters, with a mass $\sim 10^{15} h^{-1}$M$_\odot$, are expected to provide the most robust evidence of possible deviations from the standard scenario because these clusters probe the upper tail of the mass function, which is extremely sensitive to the cosmological model \citep{BBKS86,White02,Courtin11,Cui12,giocoli2018weak}, and because they are the easiest objects to detect observationally. In Sect. \ref{numRes} we showed that in the high-mass samples of simulated clusters, we observe the largest discrepancy between the values of $\varsigma$ in the two models of gravity (Table \ref{ptmin_samples}). Nevertheless, the larger sizes of the low- and intermediate-mass samples, compared to the high-mass sample, enable the best distinction between the two models of gravity (Sect. \ref{numRes3}). On the one hand, large simulated volumes can therefore provide samples with an adequate size for a solid test bench based on which the performance of $\varsigma$ as a gravitational probe can be investigated. On the other hand, low- and intermediate-mass clusters, with $\sim 1-5\cdot 10^{14} h^{-1}$M$_\odot$, can already provide observed samples that are sufficiently large to yield tight constraints on the theory of gravity. 

\section{Conclusion}\label{conclusion}

We performed the first assessment of the ability of the radial location, $\varsigma$, of the minimum of the differential radial mass profiles, $M^\prime(r)$, of clusters of galaxies to distinguish between two different models of gravity, General Relativity, with the $\Lambda$CDM model, and $f(R)$. We investigated the behavior of $\varsigma$ in a wide range of masses, $M_{200c}\sim (0.9-11)\cdot 10^{14} h^{-1}$M$_\odot$, and redshift, $0.0\leq z \leq 1.0$.  We explored the dependence of $\varsigma$ on redshift and MAR in the two models. We also showed that $\varsigma$ has a {similar} redshift, mass, and MAR dependence as the splashback radius. 

We used the DUSTGRAIN N-body simulations \citep{giocoli2018weak}. We considered the $\Lambda$CDM run and the Hu-Sawicki $f(R)$ \citep{huSaw2007} run with scalaron $f_R=-1\cdot 10^{-4}$ ($fR4$). By means of $\sim 21,000$ and $\sim 28,000$ dark matter halos in $\Lambda$CDM and $fR4$, respectively, arranged into three mass bins and 11 redshift bins, we showed that in both models $\varsigma$ decreases with increasing redshift at fixed mass and decreases with increasing mass (or equivalently, MAR) at fixed redshift. Specifically, in the $\Lambda$CDM model, $\varsigma$ decreases from $\sim 3R_{200c}$ to $\sim 2R_{200c}$ when $z$ increases from 0 to $1$. At $z\sim 0.1$, $\varsigma$ decreases from $2.8R_{200c}$ to $\sim 2.5R_{200c}$ when the MAR increases from $\sim 10^4h^{-1}$M$_\odot$~yr$^{-1}$ to  $\sim 2\times 10^5h^{-1}$M$_\odot$~yr$^{-1}$. We computed the MAR with the recipe presented in \citet{pizzardo2020,pizzardo22}, which, we demonstrated, agrees with results based on the merger trees of dark matter halos \citep{Diemer2017sparta2}. The behavior of $\varsigma$ in $fR4$ is similar to that of the $\Lambda$CDM, but at low redshifts, $\varsigma$ in $fR4$ is $\sim 15\%$ larger on average than in $\Lambda$CDM, whereas at higher redshifts ($z\sim 0.8-1.0$) the $\varsigma$s of the two models overlap. In $\Lambda$CDM, the correlations of $\varsigma$ with redshift and MAR agree with the correlations expected for the splashback radius \citep{Diemer2014,Adhikari2014,Deason21,oNeil21,Baxter21,Dacunha22}.

By using the median test, we find that when the dark matter halos with mass $\sim  10^{14} h^{-1}$M$_\odot$ at $z\leq 0.8$ are considered, $\varsigma$ can distinguish between the two theories of gravity with high significance, with a $p$-value $\lesssim 10^{-5}$. At $z\leq 0.5$, the same halos give a $p$-value $\lesssim 10^{-15}$. At the highest redshift, $\Lambda$CDM and $fR4$  are degenerate, and $\varsigma$ cannot distinguish between the two models.
Conversely, the number of halos with mass $\sim  10^{15} h^{-1}$M$_\odot$ is too small to distinguish between the two models at any redshift: 
We showed that obtaining a $p$-value $\lesssim 10^{-5}$ requires $\gtrsim 400$ halos.

Developing the full potential of the $\varsigma$ radius as a gravity probe thus requires further observational and computational efforts. 
New facilities such as the Vera Rubin Observatory \citep{Ivezic19} and Euclid \citep{Laureijs11,Sartoris16} will provide extended weak-lensing mass profiles for thousands of clusters up to $z\sim 2$.
Large dense surveys obtained with advanced spectrographs on large telescopes such as WEAVE \citep{Balcells10,Dalton12,Dalton14,Dalton16}, PFS \citep{Tamura16}, and eventually, MSE \citep{MSE19} will provide dense spectroscopic surveys of a tremendous number of clusters up to $z\sim 0.5$  that will complement the photometry. Weak gravitational lensing mass profiles estimated with sophisticated mass reconstruction methods \citep[e.g.,][]{Umetsu11,Umetsu13} combined with the caustic mass profiles \citep{Diaferio1997,Diaferio99,Serra2011,Pizzardo23} will improve the accuracy and the precision of the mass profiles at large distance from the cluster center. Larger simulated volumes will allow the identification of larger sets of the most massive halos, paving the way to tighter tests of structure formation models. 

\begin{acknowledgements}

We sincerely thank the referee for her/his insightful comments that greatly improved the quality of the paper.
We thank Marco Baldi for providing snapshots of the $\Lambda$CDM and $fR4$ runs of the DUSTGRAIN $N$-body simulation. We thank Margaret Geller, Jubee Sohn, and Benedikt Diemer for numerous stimulating discussions.
The graduate-student fellowship of M.P. was supported by the Italian Ministry of Education, University and Research (MIUR) under the Departments of Excellence grant L.232/2016. M.P. acknowledges the support of the Canada Research Chair Program and the Natural Sciences and Engineering Research Council of Canada (NSERC, funding reference number RGPIN-2018-05425). We also acknowledge partial support from the INFN grant InDark. This research has made use of NASA’s Astrophysics Data System Bibliographic Services.
\end{acknowledgements}

\bibliographystyle{aa}
\bibliography{michele}

\begin{appendix}

\section{The $f(R)$ theory of gravity}\label{append}
In the $f(R)$ model of gravity, the action $S$ of General Relativity is modified by a scalar function, $f(R)$, of the Ricci scalar curvature, $R$, as follows \citep{Buchdahl1970}:
\begin{equation}
S=\int \rm{d}^4 x \sqrt{-g}\, \left( \frac{\mathit{R + f (R)}}{16\pi \mathit{G}} +\mathcal{L}_m \right),
\end{equation}
where $g$ is the determinant of the metric, $G$ is the universal gravitational constant, and $\mathcal{L}_m$ is the matter Lagrangian. \citet{huSaw2007} proposed for $f(R)$ the form
\begin{equation}\label{f_hu}
f(R) = -m^2 \frac{c_1 \left(\frac{R}{m^2}\right)^n}{c_2 \left(\frac{R}{m^2}\right)^n +1 },
\end{equation}
where $c_1$, $c_2$, and $n$ are non-negative constant free parameters of the model, and $m^2$ is a mass scale, defined as $m^2\equiv \frac{8\pi G}{3} \rho_{M,0}= H_0^2 \frac{\rho_{M,0}}{\rho_{crit,0}}= H_0^2 \Omega_M$. This specific form of $f(R)$ guarantees (i) that at the recombination time, the model mimics $\Lambda$CDM, that is, $\lim_{R\to 0} f(R)=0$; (ii) that at the late stages of cosmic evolution, the model encompasses cosmic acceleration through the presence of a cosmological constant-like quantity, namely $\lim_{R\to \infty}f(R)={\mathrm {const}} $; and (iii) the consistence with a large plethora of low-redshift phenomena through the tuning of the three free parameters. 

In the \citet{huSaw2007} class of $f(R)$ models, any departure from $\Lambda$CDM can only arise at low redshifts, or equivalently, in the high-curvature regime. This therefore is the only relevant limit for a comparison with $\Lambda$CDM. 
In the high-curvature limit, where $c_2 \left(R \over m^2 \right)^n \rightarrow \infty$, we can expand Eq. (\ref{f_hu}) to first order in $(R /m^2 )$ as
\begin{equation}\label{h-c_approx}
f(R) \approx -m^2\frac{c_1}{c_2} + m^2 \frac{c_1}{c_2^2} \left( \frac{R}{m^2}\right)^{-n}.
\end{equation} 
In this limit, if $c_1/c_2^2 \rightarrow 0$ at fixed $c_1/c_2$, we see that $f\approx {\mathrm {const}}$, resembling the onset of a cosmological constant term. 

Using Eq. (\ref{h-c_approx}), it is customary to introduce $f_R$,
\begin{equation}\label{f_R}
f_R = \frac{d f}{d R} = -n \frac{c_1}{c_2^2} \left(\frac{R}{m^2}\right)^{-n-1},
\end{equation}
a quantity that plays a central role in the description of the evolution of the space-time and of the dynamics of the matter fields in the $f(R)$ model. $f_R$ is useful for a simple characterization of the model: From Eqs. (\ref{h-c_approx}) and (\ref{f_R}), for instance, we see that larger $n$s mimic $\Lambda$CDM until later times, whereas $f(R)$ and $\Lambda$CDM become increasingly similar as $c_1/ c_2^2$ decreases.

Equation (\ref{f_hu}) covers a wide variety of possible behaviors, which can be much different from $\Lambda$CDM. We are interested in obtaining realistic observations with small deviations from $\Lambda$CDM, however. We can pursue this aim by imposing conditions on the free parameters, in particular, by fixing a relation between $c_1$ and $c_2$. 

Following \citet{huSaw2007}, we start with the variation in the action $S$ with respect to the metric to obtain the field equations
\begin{equation}\label{eom}
G_{\mu\nu}+f_R R_{\mu\nu}-\left(\frac{f}{2}-\square f_R \right) g_{\mu\nu} - \nabla_\mu \nabla_\nu f_R = 8\pi G T_{\mu\nu},
\end{equation}
where $G_{\mu\nu}= R_{\mu\nu} -\frac{1}{2} R$, $\nabla_\mu$ is the covariant derivative, and $\square$ is the d'Alembert operator. The trace of Eq. (\ref{eom}) is
\begin{equation}\label{f_R_eq}
3\square f_R - R +f_R R -2 f = - 8\pi G \rho,
\end{equation}
which is the equation of motion of the field $f_R$. If $\square f_R$ is interpreted as a function of an effective potential $V_{\mathrm {eff}}$,
\begin{equation}
\square f_R = \frac{\partial V_{\mathrm {eff}}}{\partial f_R},
\end{equation}
Eq. (\ref{f_R_eq}) yields
\begin{equation}
\frac{\partial V_{\mathrm {eff}}}{\partial f_R} = \frac{1}{3} \left( R -f_R R +2 f - 8\pi G \rho \right),
\end{equation}
which states that $V_{\mathrm {eff}}$ has an extremum at
\begin{equation}\label{min_veff}
R -f_R R +2 f - 8\pi G \rho = 0. 
\end{equation}
We note that in the absence of any modification due to $f(R)$, Eq. (\ref{min_veff}) reduces to the standard form of the Ricci scalar,
\begin{equation}\label{gr}
R = 8\pi G \rho.
\end{equation} 

At late times, when $R/m^2 \gg 1$, Eq. (\ref{h-c_approx}) shows that the modification of gravity is approximately constant, holding $f(R)\approx -m^2 \frac{c_1}{c_2}$, because the second term on the right-hand side subleads in $(m^2 / R )$. Analogously, we  note from Eq. (\ref{f_R}) that $f_R R \propto (R/m^2)^{-n} \rightarrow
 0$. This means that when it is approximated at leading order, Eq. (\ref{min_veff}) takes the following form:
\begin{equation}\label{Rfr}
R = 8\pi G\rho - 2f \approx 8\pi G\rho +2\frac{c_1}{c_2} m^2.
\end{equation}  
In $\Lambda$CDM, when we consider the Friedmann equations with the explicit presence of the cosmological constant $\Lambda c^2 = 8\pi G \rho_\Lambda$, the $\Lambda$CDM Ricci scalar (Eq. \ref{gr}) can be rewritten in the form
\begin{equation}\label{gr1}
R = 8\pi G (\rho_M + 4\rho_\Lambda). 
\end{equation} 
Small deviations from the $\Lambda$CDM expectations are achieved when we impose equality between Eqs. (\ref{Rfr}) and (\ref{gr1}) because in this way, any discrepancy is generated only by the subleading terms of $f(R)$. This equality translates into the equality between the two second terms on the right-hand side of Eqs. (\ref{Rfr}) and (\ref{gr1}) when we set $\rho\equiv \rho_M$ in Eq. (\ref{Rfr}). We obtain the following constraint between $c_1$ and $c_2$:
\begin{equation}\label{c1ovc2}
\frac{c_1}{c_2} = 6 \frac{\Omega_\Lambda}{\Omega_M}.
\end{equation}
By taking this contraint into account and by considering a pure stiff-matter model, we can finally write Eq. (\ref{Rfr}) as
\begin{equation}
R \approx 8\pi G \rho_M + 12 \frac{\Omega_\Lambda}{\Omega_M}m^2 = 3 m^2 \left( a^{-3} +4 \frac{\Omega_\Lambda}{\Omega_M} \right),
\end{equation} 
whose current value is 
\begin{equation}\label{R0}
R_0 = 3m^2\left(1+4\frac{1-\Omega_M}{\Omega_M} \right) = m^2 \left( \frac{12}{\Omega_M} - 9 \right).
\end{equation}
When $n$ is fixed, the model is left with only one free parameter: $c_2$, which is more commonly rephrased in terms of the current value of $f_R$, $f_{R0}$, known as the scalaron. 
Its expression follows from Eqs. (\ref{f_R})-(\ref{c1ovc2})-(\ref{R0}). When we take $n=1$, the chosen value in the DUSTGRAIN suite, the scalaron is
\begin{equation}
f_{R0} = -\frac{1}{c_2} \frac{6\Omega_\Lambda}{\Omega_M} \left(\frac{m^2}{R_0} \right)^2. 
\end{equation}
For the purposes of our work, we considered the DUSTGRAIN $f(R)$ run with $f_{R0}=-1\cdot 10^{-4}$.

\end{appendix}

\end{document}